\documentclass[preprintnumbers,amsmath,amssymb,twocolumn]{revtex4}%

\usepackage{graphicx}
\usepackage{dcolumn}
\usepackage{bm}
\usepackage{color}

\begin{document}

\title{Quasinormal modes of test fields around regular black holes}

\author{Bobir Toshmatov$^{(1)}$}
\email{b.a.toshmatov@gmail.com}

\author{Ahmadjon Abdujabbarov$^{(1,2,3)}$}
\email{ahmadjon@astrin.uz}

\author{Zden\v{e}k Stuchl\'{i}k$^{(1)}$}
\email{zdenek.stuchlik@fpf.slu.cz}

\author{Bobomurat Ahmedov$^{(2,3)}$}
\email{ahmedov@astrin.uz}

\affiliation{$^{1}$ Faculty of Philosophy and Science, Institute of Physics, Silesian University in Opava,\\ Bezru\v{c}ovo n\'{a}m\v{e}st\'{i} 13, CZ-74601 Opava, Czech Republic\\
$^{2}$ Institute of Nuclear Physics, Ulughbek, Tashkent 100214, Uzbekistan\\
$^{3}$ Ulugh Beg Astronomical Institute, Astronomicheskaya 33, Tashkent 100052, Uzbekistan}

\begin{abstract}
We study scalar, electromagnetic and gravitational test fields in the Hayward, Bardeen and Ay\'{o}n-Beato-Garc\'{i}a regular black hole spacetimes and demonstrate that the test fields are stable in all these spacetimes.  Using the sixth order WKB approximation of the linear "axial" perturbative scheme, we determine dependence of the quasinormal mode (QNM) frequencies on the characteristic parameters of the test fields and the spacetime charge parameters of the regular black holes. We give also the greybody factors, namely the transmission and reflection coefficients of scattered scalar, electromagnetic and gravitational waves. We show that damping of the QNMs in regular black hole spacetimes is suppressed in comparison to the case of Schwarzschild black holes, and increasing charge parameter of the regular black holes increases reflection and decreases transmission factor of incident waves for each of the test fields.
\end{abstract}

\maketitle

\section{Introduction}

It is well known that in reality a black hole (BH) can never be in an isolated state. It always interacts with  matter and fields around it and as the result of these interactions it takes a perturbed state. Behaviour of the perturbative test scalar, electromagnetic, gravitational or Dirac fields is an interesting topic of the black hole physics, of special interest is stability of these fields treated as perturbations in linear regime.

Perturbed BH are characterized by gravitational (electromagnetic, scalar) waves that are characterized by complex frequency which is called \textit{quasinormal mode} (QNM). Real and imaginary parts of this complex frequency govern the real oscillation frequency and dissipation or enhancing of these oscillations, respectively. If we are able to detect these waves we can obtain more precise imaginations about BHs. This is why the perturbations of a BH are the most momentous topic for taking information from the BH. For these purposes, perturbations of BHs have been studied for a long time. Metric perturbations of the Schwarzschild spacetime were studied long time ago by Regge and Wheeler for "axial" (or odd parity)~\cite{Regge} and by Zerilli for "polar" (or even parity)~\cite{Zerilli} perturbations by adding small perturbation terms onto the unperturbed background, requiring that these perturbations do not change the stress energy tensor of the black hole spacetime. Later Vishveshwara studied numerically scattering of waves on the Schwarzschild BH background ~\cite{Vishveshwara} see for the review~\cite{Nagar}. Afterwards, Chandrasekhar presented monograph about perturbation theory of BHs summarizing all results that had been got by that time~\cite{Nagar}. All the authors showed that the above mentioned perturbative studies of test fields on stationary backgrounds lead to a Schr\"{o}dinger-like equation with complex frequency and a specific effective potential. This equation can be solved using semiclassical or numerical methods depending on the character of the effective potential. So far, the Schr\"{o}dinger equation has been solved for several potentials in terms of specific functions or numerically~\cite{Boonserm}. We should note that reviews of QNMs of BHs and neutron stars have been written in~\cite{Kokkotas,Berti,Konoplya0}. Kodama and Ishibashi extended the calculations of QNMs into BHs in higher dimensions in their series of papers~\cite{Meq1,Meq2,Meq3}.

One of the standard methods for the QNM calculations is the WKB (Wentzel-Kramers-Brillouin) approximative method that was applied for the first time by Schutz and Will~\cite{Schutz}. Afterwards, in order to increase accuracy of this method it was extended to the third and sixth order by Iyer and Will~\cite{Iyer} and Konoplya~\cite{WKB6}.

Up to now, QNMs of several BH spacetimes have been calculated by several methods, such as the WKB method, integration of the wavelike equations, fit and interpolation approaches, Frobenius method, method of continued fractions, the Mashhoon method and so on \cite{Leaver,Iyer1,Nollert,Motl1,Schw2,Schw3,Padmanabhan,Cardoso,Cardoso2,Cardoso3, Schw.dS,Otsuki,Moss,Choudhury,Schw.dS2,KZ,AZ,RNds,Motl2,KZ1,Kerr,Musiri,Hod,KN,Keshet,Zhu,Wang}. In~\cite{Kodama2,Bronnikov,Lemos,Abdalla,Naresh} instabilities of the BHs against to the perturbations have been studied. QNMs of regular and dirty BHs has been investigated in~\cite{RBH,Li,Medved,Nomura,Fernando}.

So far, curvature singularity has been considered to be the fundamental problem of General Relativity, as the curvature singularity cannot be explained by the General Relativity itself. As the result of efforts to achieve a new solution without singularity, in 1968 Bardeen presented a new regular BH solution of modified Einstein field equations coupled to a nonlinear electrodynamics of a self-gravitating magnetic monopole~\cite{Bardeen}. Afterwards, in 1998 Ay\'{o}n-Beato and Garc\'{i}a presented a new charged regular BH solution based on the Einstein gravity and non-linear electrodynamics. For finding this regular BH solution, as a source of charge they took the nonlinear electric field for the solution of Maxwell field equations~\cite{ABG}. All the subtleties of the non-linear electrodynamics and the presence of the regular BH solutions were discussed in the series of papers of Bronnikov~\cite{Bronnikov2,Bronnikov3,Bronnikov4}. Regular BH solutions are discussed also in the framework of the so called Modified Gravity (MOG) introduced as tensor-vector-scalar (TeVeS) gravitation theory by Moffat~\cite{Moffat}. In 2006 Hayward found new regular BH solution that is free of charge term and quiet similar to the Bardeen one in terms of physical aspects~\cite{Hayward}. Geodesic motion in the field of regular BHs has been studied in~\cite{Garcia,Patil,Bobir}. For the regular no-horizon highly curved spacetimes, being complementary to the regular BH spacetimes, the geodesic motion has been studied in~\cite{Stu-Sche:IJMPD:2015:}, the extraordinary effect of ghost images that do not occur neither in BH or naked singularity spacetimes has been exposed in~\cite{Stuchlik2}. Geodesic structure of the no-horizon spacetimes is of similar character as those of the naked singularity Reissner-Nordstr\"{o}m~\cite{SH,Pugliese1,Pugliese2,Pugliese3} or Kehagias-Sfetsos spacetimes~\cite{Vieira,Schee,Daniela}, but differs significantly for the Kerr naked singularity spacetimes~\cite{Stuchlik3,Stuchlik4,Stuchlik5}. Here we concentrate on the behaviour of the scalar, electromagnetic and gravitational perturbative fields in the regular Hayward, Bardeen and Ayon-Beato-Garcia (ABG) BH spacetimes. The dynamical stability of black hole solutions in self-gravitating nonlinear electrodynamics with respect to arbitrary linear fluctuations of the metric and electromagnetic field has been analyzed in~\cite{Moreno}. We calculate the quasinormal modes and reflection or transmission coefficients using the 6th order WKB method.

This paper is organized as follows. Section~II gives brief description of the Hayward, Bardeen and ABG regular BH spacetimes. In section~III the perturbative equations of the test scalar, electromagnetic and gravitational fields in given backgrounds are reduced to the Schr\"{o}dinger-like wave equation. Stability of the perturbative scalar, electromagnetic and gravitational fields in the regular BH spacetimes is treated in the linear regime in section~IV. Section~V is devoted to the calculations of the eikonal limit for the frequency of the QNM in the regular BH spacetimes under consideration and numerical calculations are made by the 6th order WKB approximation. In section~VI we analyze transmission and reflection of incident waves through the potential barrier given by the regular BHs. In section~VII we summarize our main results.

\section{Regular black holes}

Spacetime line element of the spherically symmetric regular BH spacetimes can be expressed in the form
\begin{eqnarray}\label{01}
ds^2=-f(r)dt^2+\frac{1}{f(r)}dr^2+r^2d\theta^2+r^2\sin^2\theta d\varphi^2\ ,
\end{eqnarray}
with the lapse function $f(r)$ determined by the formula
\begin{eqnarray}\label{02}
f_i(r)=1-\frac{2m_i(r)}{r}\ ,
\end{eqnarray}
where index $i=1,2,3$ corresponds to the Hayward, Bardeen and ABG BHs, respectively.

For the Hayward BH~\cite{Hayward}
\begin{eqnarray}\label{02}
m_1(r)=\frac{Mr^3}{r^3+Q_1^3},
\end{eqnarray}
for the Bardeen BH~\cite{Bardeen}
\begin{eqnarray}\label{03}
m_2(r)=\frac{Mr^3}{(r^2+Q_2^2)^{3/2}},
\end{eqnarray}
for the ABG BH~\cite{ABG}
\begin{eqnarray}\label{04}
m_3(r)=\frac{Mr^3}{(r^2+Q_3^2)^{3/2}}-\frac{Q_3^2r^3}{2(r^2+Q_3^2)^{2}}.
\end{eqnarray}
Here $M$, $Q_2$ and $Q_3$ are mass, magnetic and electric (or magnetic) charge, respectively, while $Q_1$ is some real positive constant.

The Hayward and Bardeen regular BHs solutions asymptotically behave as the Schwarzschild one at large distances $r\rightarrow\infty$
\begin{eqnarray}
\lim_{r\to\infty}f_{1,2}(r)=1-\frac{2M}{r}+O\left(\frac{1}{r^4}\right)\ ,
\end{eqnarray}
The ABG regular BH solution asymptotically behaves as the Reissner-Nordstr\"{o}m one
\begin{eqnarray}
\lim_{r\to\infty}f_3(r)=1-\frac{2M}{r}+\frac{Q_{3}^2}{r^2}+O\left(\frac{1}{r^3}\right).
\end{eqnarray}
When $r\rightarrow0$, the Bardeen, Hayward and ABG BH spacetimes behave as the de Sitter spacetime
\begin{eqnarray}
\lim_{r\to0}f_{1,2}(r)=1-\frac{2M}{Q_{1,2}^3}r^2+O\left(r^5\right)\ ,\\
\lim_{r\to0}f_3(r)=1-\frac{2M}{Q_{3}^3}r^2+\frac{r^2}{Q_3^2}+O\left(r^3\right).
\end{eqnarray}
The ABG BH spacetimes may demonstrate an anti de-Sitter behaviour near the origin of coordinates, however it is related to the no-horizon case ($Q>2M$)~\cite{Stu-Sche:IJMPD:2015:}.

\begin{figure}[h]
\centering
\includegraphics[width=8cm]{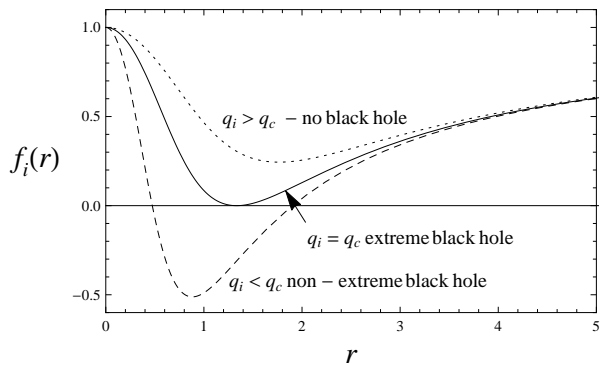}
\caption{\label{lapse} Radial dependence of lapse function $f_i(r)$ of the regular BHs for different values of the parameter $q_i$. Where $q_i$ is $g, q, d$ for the Hayward, Bardeen and ABG spacetimes, respectively.}
\end{figure}

For convenience we turn into dimensionless quantities: $r\rightarrow r/M$, $g\rightarrow Q_1/M$, $q\rightarrow Q_2/M$ and $d\rightarrow Q_3/M$. Then for the Hayward BHs
\begin{eqnarray}\label{05}
m_1(r)=\frac{r^3}{r^3+g^3},
\end{eqnarray}
for the Bardeen BHs
\begin{eqnarray}\label{06}
m_2(r)=\frac{r^3}{(r^2+q^2)^{3/2}},
\end{eqnarray}
and for the ABG BHs
\begin{eqnarray}\label{07}
m_3(r)=\frac{r^3}{(r^2+d^2)^{3/2}}-\frac{d^2r^3}{2(r^2+d^2)^{2}}.
\end{eqnarray}
The horizons of the static spherically symmetric BHs are defined by vanishing of the time metric component, $g_{tt}=0$, i.e., the lapse function $f(r,q_i)=0$. By solving this equation with respect to the radial coordinate $r$, one obtains the outer event horizon $r_+$, and the inner horizon $r_-$. One can see from the Fig.~\ref{lapse} that the regular Hayward, Bardeen and ABG spacetimes can have two horizons, corresponding to the BH spacetimes, one horizon, corresponding to the extremal BH spacetimes, and no horizon, corresponding to the no-horizon spacetimes, in dependence on the values of the mass $m$ and parameters $q_i$. In order to find the critical value of the charge $q_c$ corresponding to the extremal BH spacetimes (when $r_+=r_-$), one can use the system of equations
\begin{eqnarray}\label{09}
g_{tt}=0, \quad g_{tt,r}=0.
\end{eqnarray}
By solving above equations simultaneously, we obtain the critical charge of the Hayward spacetimes to be $g_c\approx1.0583$ while related radius of the horizon reads $r_c=1.3333$. It is shown in Fig.~\ref{lapse} that in the case $q<q_c$ there are two horizons, in the case $q>q_c$ there are no horizons. For the Bardeen and ABG spacetimes we arrive at the critical charges and related horizon radii expressed in dimensionless form: $q_c\approx0.7698$, $r_c=1.0887$ and $d_c\approx0.6342$, $r_c=1.005$, respectively.

\section{Equations governing test fields}

Let us consider a massless scalar field $\Phi$ in the spherical symmetric regular BH spacetimes, obeying the Klein-Gordon equation~\cite{Moreno}
\begin{eqnarray}\label{14}
\Box\Phi\equiv\frac{1}{\sqrt{-g}}\partial_{\mu}\left(\sqrt{-g}g^{\mu\nu}\partial_{\nu}\Phi\right)=0\ ,
\end{eqnarray}
where $\partial_{\mu}$ is the partial derivative, and $-g$ is the absolute value of the determinant of the spacetime metric~(\ref{01}).

In order to reduce the Klein-Gordon equation~(\ref{14}) into the two dimensional wave equation, we assume the scalar field separated in the standard form
\begin{eqnarray}\label{15}
\Phi(t,r,\theta,\phi)=\frac{1}{r}R(r,t)Y_{\ell}^m(\theta,\phi)\ ,
\end{eqnarray}
where $Y_{\ell}(\theta,\phi)$ is the so called spherical harmonic function of degree $\ell$ related to the angular coordinates $\theta, \phi$, and it fulfills the relations
\begin{eqnarray}\label{15.1}
\nabla_{\theta,\phi}^2Y_{\ell}^m(\theta,\phi)=&&\left[\frac{1}{\sin\theta}\partial_{\theta}\left(\sin\theta\partial_{\theta}\right)+ \frac{1}{\sin^2\theta}\partial_{\phi\phi}^2\right]Y_{\ell}^m(\theta,\phi)\nonumber\\
&&=-\ell(\ell+1)Y_{\ell}^m(\theta,\phi)\ .
\end{eqnarray}
Inserting the separated scalar function~(\ref{15}) into~(\ref{14}), we obtain the Regge-Wheeler wave equation~\cite{Regge} relating the time and radial dependence of the scalar function
\begin{eqnarray}\label{16}
\left[\frac{\partial^2}{\partial t^2}-\frac{\partial^2}{\partial r_\ast^2}+V_s(r)\right]R(r,t)=0.
\end{eqnarray}
In the Regge-Wheeler equation, $r_\ast$ is the so called "tortoise" radial coordinate defined by
\begin{eqnarray}\label{17}
r_\ast=\int\frac{dr}{f(r)}.
\end{eqnarray}
Since $f(r)$ is the lapse function governing the event horizon location, $r_\ast$ approaches $-\infty$ as $r$ approaches the event horizon of the BH from infinity, thus
\begin{eqnarray}\label{17.1}
r_\ast\in(-\infty, +\infty) \quad , for \quad r\in(r_+, +\infty).
\end{eqnarray}
Therefore, the Regge-Wheeler wave-like equation~(\ref{16}) will be restricted here only to the regions located outside the event horizon, $r>r_+$.

The effective potential $V(r)$ in the expression~(\ref{16}) depends on the specific field under consideration. For the massless scalar fields it reads
\begin{eqnarray}\label{16.1}
V_s(r)=f(r)\left[\frac{\ell(\ell+1)}{r^2}+\frac{1}{r}\frac{d f(r)}{dr}\right],
\end{eqnarray}
where parameter $\ell$ related to the spherical harmonics represents the orbital angular momentum and takes only nonnegative integers. Generalized form of the effective potential to higher spin (bosonic) fields can be written in the form~\cite{Medved,Nomura}
\begin{eqnarray}\label{18}
V(r)=f(r)\left[\frac{\ell(\ell+1)}{r^2}+(1-s^2)\frac{2m(r)}{r^3}\right.\nonumber\\
\left.+(1-s)\left(\frac{1}{r}\frac{d f(r)}{dr}-\frac{2m(r)}{r^3}\right)\right]\ ,
\end{eqnarray}
where the multipole number $\ell$ is restricted by $\ell\geq s$, and $s$ is the spin of the perturbative field
\begin{eqnarray}\label{value_s}
s=\left\{
\begin{array}{rc}
   0, \quad & scalar\ perturbation, \\
   1, \quad & electromagnetic\ perturbation, \\
   2, \quad & gravitational\ perturbation.
\end{array}
\right.
\end{eqnarray}
\begin{figure}[h]
\centering
\includegraphics[width=8cm]{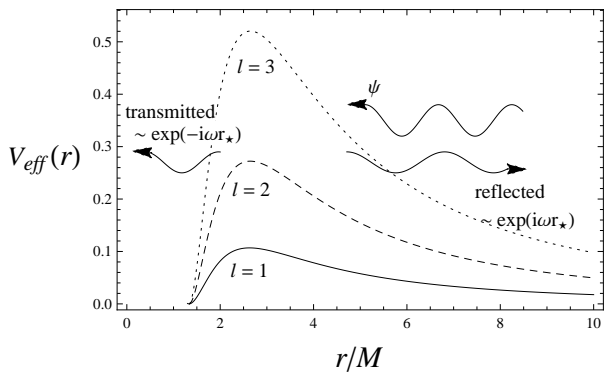}
\caption{\label{veff} Dependence of $V$ on radial coordinate $r$ for several values of orbital quantum number $\ell$.}
\end{figure}
One can deduce from Fig.~\ref{veff} that the angular momentum parameter $\ell$ increases the height of the potential barrier governed by the effective potential.

In order to solve the Regge-Wheeler differential equation~(\ref{16}) we separate the $r,t$ variables in the wave function assuming
\begin{eqnarray}\label{20}
R(r,t)=\psi(r)\xi(t)\ .
\end{eqnarray}
Putting the separated function~(\ref{20}) into~(\ref{16}), using the stationarity of the spacetime, and considering a wave incoming from infinity towards the BH,
\begin{eqnarray}\label{21}
\xi(t)\sim e^{-i\omega t},
\end{eqnarray}
we obtain the radial part of the Regge-Wheeler wave equation in the form of the standard Schr\"{o}dinger  equation
\begin{eqnarray}\label{wave_equation}
\frac{d^2\psi(r)}{d r_\ast^2}+\left[\omega^2-V(r)\right]\psi(r)=0\ .
\end{eqnarray}
As mentioned before, the QNM frequency can be generally expressed in the form
\begin{eqnarray}\label{22}
\omega=Re(\omega)+i Im(\omega)=\omega_R+i\omega_I
\end{eqnarray}
where $\omega_R$ is the real oscillation frequency, and $\omega_I$ is the imaginary oscillation frequency that represents damping or enhancing of the oscillatory mode.

\section{Stability of regular BH spacetimes}

Equation~(\ref{wave_equation}) is treated considering that there are only outgoing waves at infinity and only ingoing waves at the event horizon of regular BHs.

Let us rewrite the wavelike equation~(\ref{wave_equation}) in the form
\begin{eqnarray}\label{wave_equation2}
-\frac{d^2\psi(r)}{d r_\ast^2}+V(r)\psi(r)-\omega^2\psi(r)=0\ .
\end{eqnarray}
In order to study stability of the regular BHs against "axial" perturbations (or, more precisely, stability of the perturbative fields in the given spacetimes), we multiply both sides of equation~(\ref{wave_equation2}) by complex conjugate of the wave function $\psi(r)$
\begin{eqnarray}\label{wave_equation3}
-\psi^\ast(r)\frac{d^2\psi(r)}{d r_\ast^2}+V(r)\psi(r)\psi^\ast(r)-\omega^2\psi(r)\psi^\ast(r)=0\ .\nonumber\\
\end{eqnarray}
Integrating equation~(\ref{wave_equation3}) along whole the range of the "tortoise" coordinate $r_\ast$, we arrive to the formula
\begin{eqnarray}\label{wave_integral}
\int_{-\infty}^{+\infty}\left[V(r)\left|\psi(r)\right|^2+ \left|\psi'(r)\right|^2\right]dr_\ast- \psi^\ast(r)\psi'(r)|_{-\infty}^{+\infty}\nonumber\\
-\omega^2\int_{-\infty}^{+\infty}\left|\psi(r)\right|^2dr_\ast=0\ ,
\end{eqnarray}
where $" ' "$ stands for derivative with respect to the "tortoise" coordinate $r_\ast$.

As mentioned in Sec.~III, the effective potential related to the Hayward, Bardeen and ABG regular BHs,  $V(r)$, is always positive in the region $r_\ast\in(-\infty,+\infty)$, i.e., outside the event horizon $r\in[r_+,+\infty)$. Now we rewrite the equation~(\ref{wave_integral}) for the imaginary part using the boundary conditions~(\ref{boundary1}),~(\ref{boundary2}) and positivity of the effective potential outside the horizon~\cite{Kimura}
\begin{eqnarray}\label{wave_integral1}
\omega_R\left[\left|\psi(+\infty)\right|^2+\left|\psi(-\infty)\right|^2+ 2\omega_I\int_{-\infty}^{+\infty}\left|\psi(r)\right|^2dr_\ast\right]=0.\nonumber\\
\end{eqnarray}
According to this relation we can deduce that positivity of the real part of the frequency, $\omega_R>0$, requires the imaginary part of the frequency to be negative, $\omega_I<0$. This means that the perturbative test fields have no exponentially growing modes and the regular BH spacetimes are stable under linear odd parity perturbations.

\section{WKB method and numerical results}

\textit{\textbf{Boundary conditions.}} Considering the boundary condition for an incoming wave at the event horizon, one can see from Fig.~\ref{veff} that for $r\rightarrow r_+$, the effective potential $V_{eff}\rightarrow0$, therefore, in this limit, the incoming wave function $\psi(r)$ behaves as
\begin{eqnarray}\label{boundary1}
\psi(r)\sim e^{-i\omega r_\ast} \quad as \quad r\rightarrow r_+ \quad (r_\ast\rightarrow-\infty).
\end{eqnarray}
Considering the boundary condition for an outgoing wave at infinity we have to use the fact that all the considered regular BH spacetimes are asymptotically flat, i.e., at infinity they tend to the Minkowski, flat spacetime. In the Minkowski spacetime, the effective potential vanishes, $V_{eff}=0$. Therefore one can write the boundary condition for the outgoing wave as
\begin{eqnarray}\label{boundary2}
\psi(r)\sim e^{i\omega r_\ast} \quad as \quad r\rightarrow \infty \quad (r_\ast\rightarrow+\infty).
\end{eqnarray}

\textit{\textbf{WKB method.}} So far, the Schr\"{o}dinger equation has been solved for several (solvable) potentials in terms of special functions or numerically. The general approach to solve the Schr\"{o}dinger equation for properly defined effective potentials is to reduce this equation to the equation for the  hypergeometric functions or some other special functions.

One of the methods of solving the Schr\"{o}dinger equation is the WKB one. This method was applied for the QNMs of test field in the field of BHs for the first time by Schutz~\cite{Schutz}. The WKB method is a semi-analytic technique used to solve Schr\"{o}dinger-type equations such as~(\ref{wave_equation}). It is used mostly for the time-independent case, in other words, for an eigenstate of QNM having two turning points in the effective potential. However, this method also have its validity region. WKB has very good accuracy, if the two turning points (solutions of equation~$\omega^2-V(r)=0$) are very close to each other. In other words, in the case when $[V(r)-\omega^2]_{max}\ll[\omega^2-V(\pm\infty)]$. In the WKB method, the total energy, $\omega^2-V(r)$, is expanded to the Taylor series near the maximum of the effective potential nearby the turning points. The WKB method has been extended to the third and the sixth order by Iyer and Will~\cite{Iyer} and Konoplya~\cite{WKB6}, respectively.

In this paper we use the sixth order WKB method of calculation of the quasi-normal modes~\cite{WKB6} that is governed by the relation
\begin{eqnarray}\label{22}
\frac{i(\omega^2-V(r_0))}{\sqrt{-2V''(r_0)}}+\sum_{i=2}^{6}\Lambda_i=n+\frac{1}{2}\ ,
\end{eqnarray}
where $V''(r_0)$ is the value of the second derivative of the effective potential with respect to $r$ at its maximum point $r_0$ defined by the solution of the equation $dV/dr_\ast|_{r_\ast=r_0}=0$. $\Lambda_i$ are constants coming from the second up to the sixth order WKB corrections~\cite{Iyer,WKB6}.

\textit{\textbf{The eikonal limit}}. It has been already shown in~\cite{Kokkotas},~\cite{Konoplya0} that the WKB method works with large accuracy for large values of the multipole quantum number $\ell$. For $\ell\gg1$, frequency of QNMs of a massless "axial" perturbation field with arbitrary spin $s$ can be found analytically by using the first order WKB approach. In the case of the Hayward, Bardeen and ABG regular BH spacetimes we arrive to the formulae
\begin{eqnarray}\label{largel1}
\omega_H=\sqrt{\frac{g^3+r_0^2(r_0-2)}{r_0^2(g^3+r_0^3)}}\left[\ell+\frac{1}{2}-i C_H(n+\frac{1}{2})\right]\nonumber\\
+ O\left(\frac{1}{\ell}\right)\ ,
\end{eqnarray}
\begin{eqnarray}\label{largel2}
\omega_B=\sqrt{\frac{(q^2+r_0^2)^{3/2}-2r_0^2}{r_0^2(q^2+r_0^2)^{3/2}}}\left[\ell+\frac{1}{2}-i C_B(n+\frac{1}{2})\right]\nonumber\\
+ O\left(\frac{1}{\ell}\right)\ ,
\end{eqnarray}
\begin{eqnarray}\label{largel3}
\omega_{ABG}=\sqrt{\frac{\left(d^2+r_0^2\right)^{2}+r_0^2\left(d^2-2\sqrt{d^2+r_0^2}\right)}{r_0^2\left(d^2+r_0^2\right)^{2}}}\nonumber\\ \times\left[\ell+\frac{1}{2}-i C_{ABG}(n+\frac{1}{2})\right]+ O\left(\frac{1}{\ell}\right)\ ,
\end{eqnarray}
Interestingly, the zeroth order term of the quasinormal frequency $\omega$, expressed in powers of $1/\ell$, does not depend on the spin $s$. Therefore, in any regular BH spacetime oscillations of the quasinormal modes with large values of the multipole quantum number $\ell$ have the same frequency for each "axial" perturbative test field, independent of the spin of the field.

The cumbersome constants $C_B$, $C_H$ and $C_{ABG}$ are functions of the dimensionless parameters $g$, magnetic charge $q$ and electric charge $d$, respectively. In the vacuum cases ($g=0$, $q=0$, $d=0$), there is $C_H=1$, $C_B=1$, $C_{ABG}=1$, and the above given expressions are identical with the one corresponding to the Schwarzschild BH~\cite{Konoplya1}.

For the extremal Hayward BH spacetimes, the effective potential reaches its maximum at $r_0\approx 2.6524$ and the QNM frequency takes the form
\begin{eqnarray}\label{limit1}
\lim_{\ell\to\infty} \omega_H\approx0.2034\left[\ell+\frac{1}{2}-i 0.8011(n+\frac{1}{2})\right]\ .
\end{eqnarray}
In the extremal Bardeen BH spacetimes, there is $r_0\approx 2.3012$ and the QNM frequency takes the form
\begin{eqnarray}\label{limit2}
\lim_{\ell\to\infty} \omega_B\approx0.2210\left[\ell+\frac{1}{2}-i 0.7048(n+\frac{1}{2})\right]\ .
\end{eqnarray}
In the extremal ABG BH spacetimes, there is $r_0\approx 2.1372$ and the QNM frequency takes the form
\begin{eqnarray}\label{limit3}
\lim_{\ell\to\infty} \omega_{ABG}\approx0.2339\left[\ell+\frac{1}{2}-i 0.6854(n+\frac{1}{2})\right]\ .
\end{eqnarray}

The Schr\"{o}dinger-like wave equation~(\ref{wave_equation}) with the effective potential~(\ref{18}) containing the lapse function $f(r)$ related to the regular BHs is not solvable analytically. Using the sixth order WKB method, we calculate numerically the frequency of QNMs in regular BHs for the scalar, electromagnetic and gravitational perturbative fields. It is known that the WKB method demonstrates high accuracy for low overtones with small imaginary part of the QNM frequency if $\ell\geq n$~\cite{Konoplya12}.

\begin{figure*}[t!.]
\begin{center}
\includegraphics[width=0.48\linewidth]{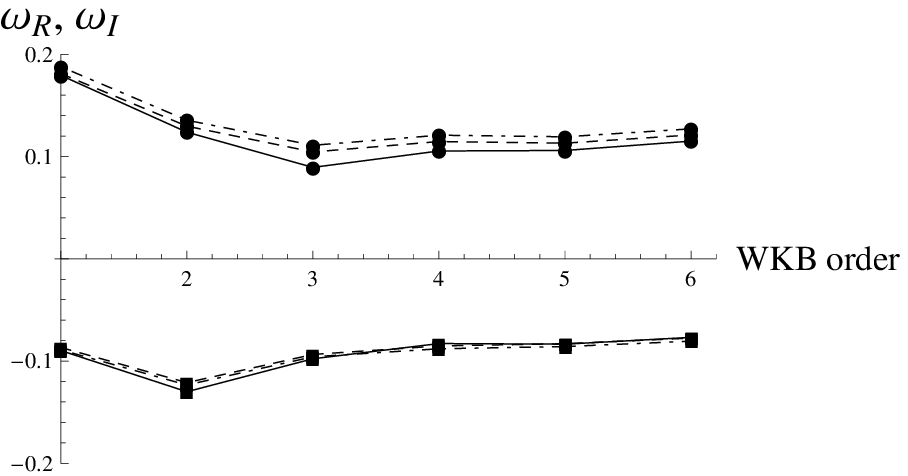}
\includegraphics[width=0.48\linewidth]{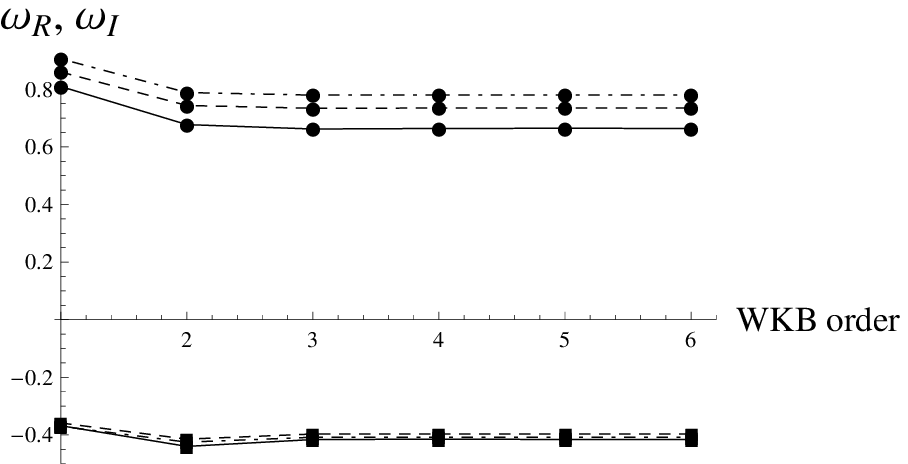}
\end{center}
\caption{\label{wkbscalar} Real $\omega_R$ (with $\bullet$) and Imaginary $\omega_I$ (with $\blacksquare$) parts of the QNM frequency as a function of WKB order for scalar perturbations ($s=0$) of the extremal Hayward (solid), extremal Bardeen (dashed) and extremal ABG (dotdashed) BHs are given for the $l=0$, $n=0$ (left) and $l=3$, $n=2$ (right) modes.}
\end{figure*}

\begin{figure*}[t!.]
\begin{center}
\includegraphics[width=0.48\linewidth]{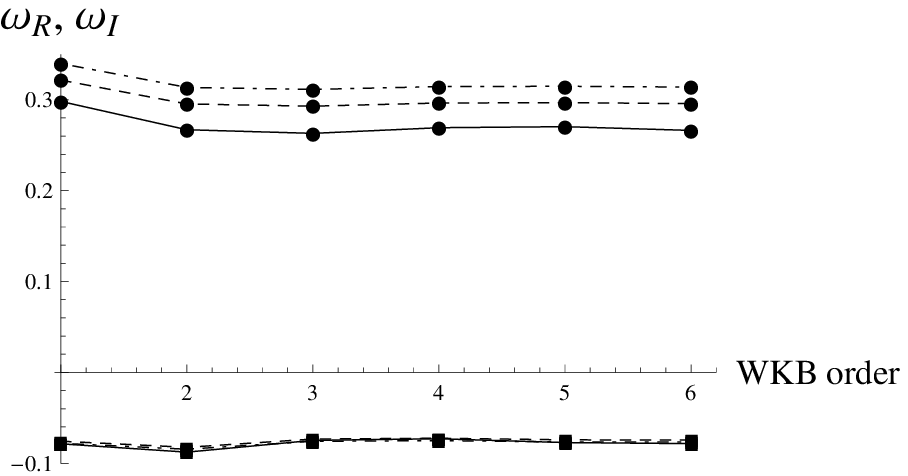}
\includegraphics[width=0.48\linewidth]{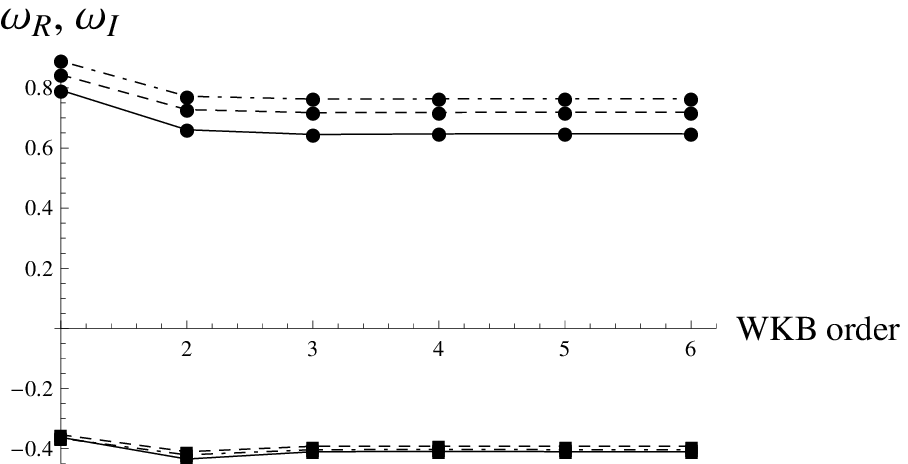}
\end{center}
\caption{\label{wkbem} Real $\omega_R$ (with $\bullet$) and Imaginary $\omega_I$ (with $\blacksquare$) parts of the QNM frequency as a function of the WKB order for the electromagnetic perturbative fields ($s=1$) in the extremal Hayward (solid), extremal Bardeen (dashed) and extremal ABG (dotdashed) BHs are given for the $l=1$, $n=0$ (left) and $l=3$, $n=2$ (right) modes.}
\end{figure*}

\begin{figure*}[t!.]
\begin{center}
\includegraphics[width=0.48\linewidth]{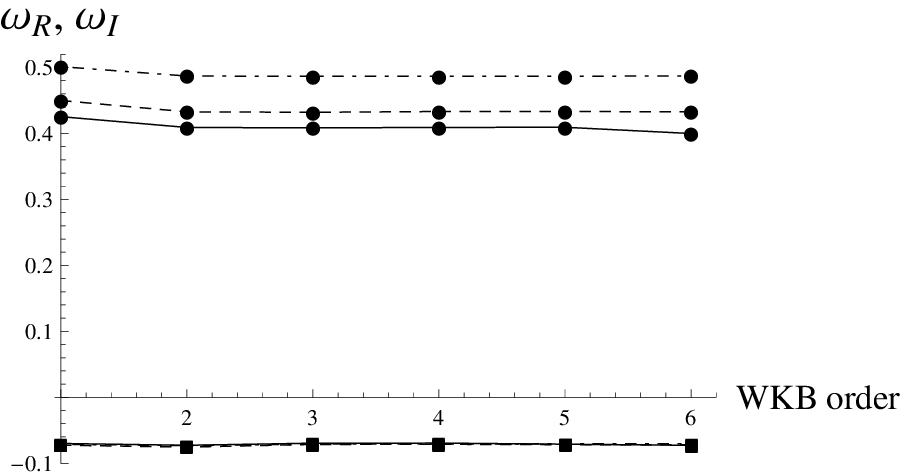}
\includegraphics[width=0.48\linewidth]{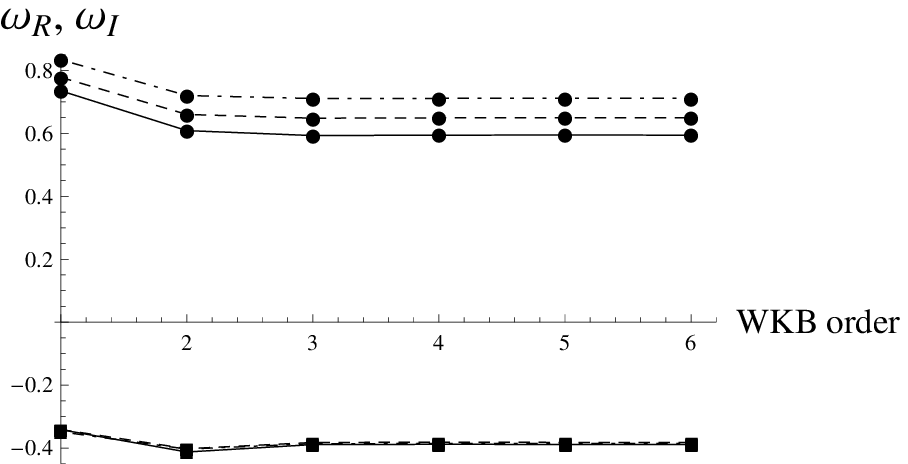}
\end{center}
\caption{\label{wkbgraviton}Real $\omega_R$ (with $\bullet$) and Imaginary $\omega_I$ (with $\blacksquare$) parts of the QNM frequency are given as a function of the WKB order for the gravitional perturbative fields ($s=2$) in the extremal Hayward (solid), extremal Bardeen (dashed) and extremal ABG (dotdashed) BHs for the $l=2$, $n=0$ (left) and $l=3$, $n=2$ (right) modes.}
\end{figure*}

In Figs~\ref{wkbscalar} -~\ref{wkbgraviton} the real and imaginary parts of the QNM frequencies are presented for the Hayward, Bardeen and ABG regular BHs being given as a function of the WKB order.

\begin{figure*}[t!.]
\begin{center}
\includegraphics[width=0.32\linewidth]{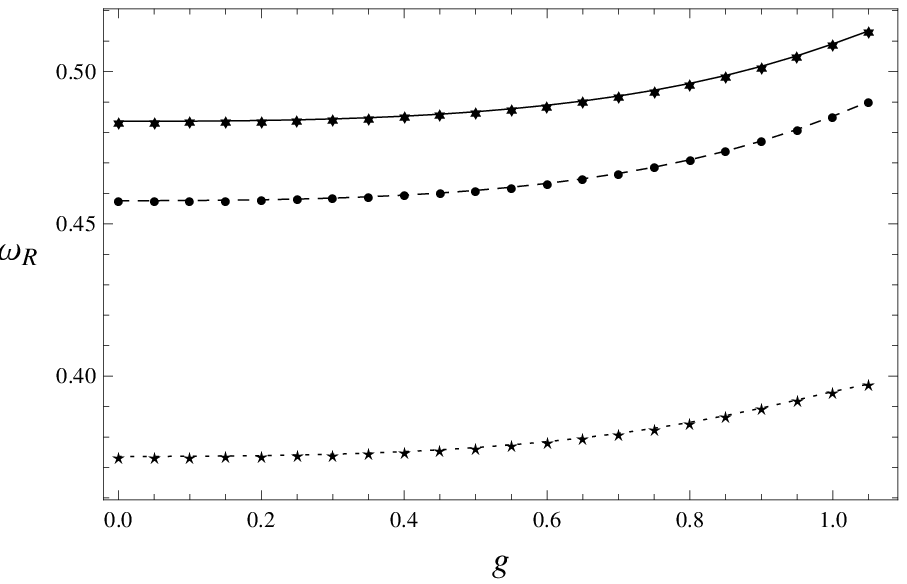}
\includegraphics[width=0.32\linewidth]{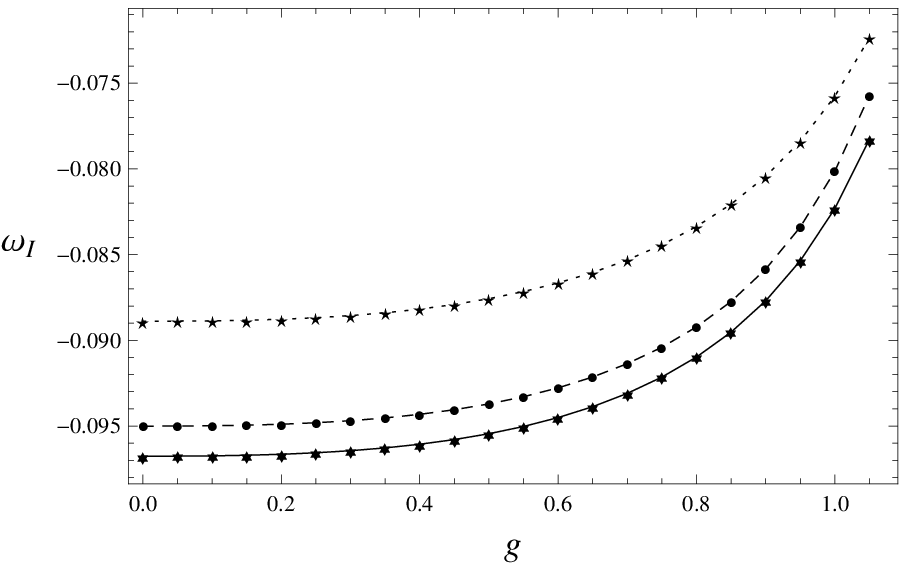}
\includegraphics[width=0.32\linewidth]{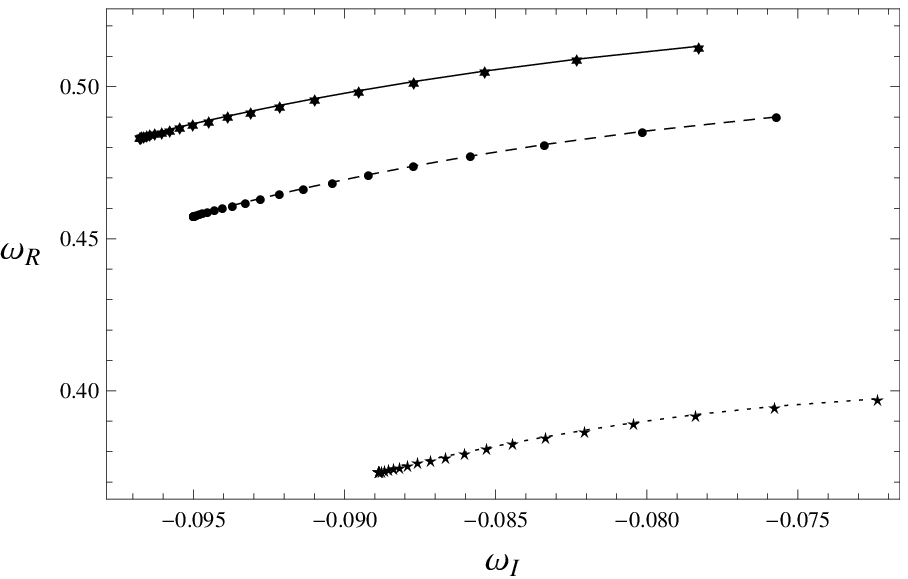}
\end{center}
\caption{\label{wkbwq} Dependence of the real (left) $\omega_R$ and imaginary (right) $\omega_I$ part of the QNM frequencies of the scalar (solid $\ast$), electromagnetic (dashed $\bullet$) and gravitational (dotted $\star$) perturbative fields on the charge parameter $g$ of the Hayward BHs is given for the $l=2$, $n=0$ mode.}
\end{figure*}

\begin{figure*}[t!.]
\begin{center}
\includegraphics[width=0.32\linewidth]{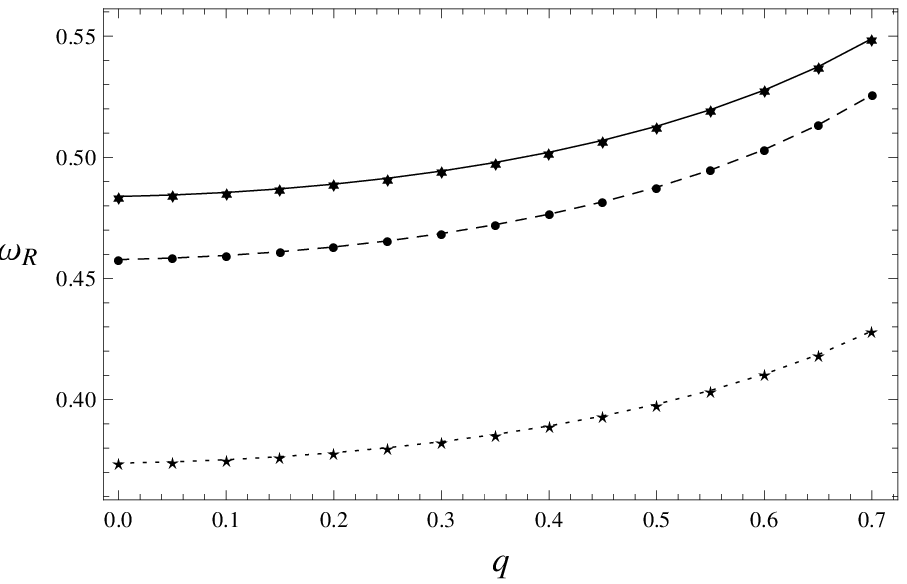}
\includegraphics[width=0.32\linewidth]{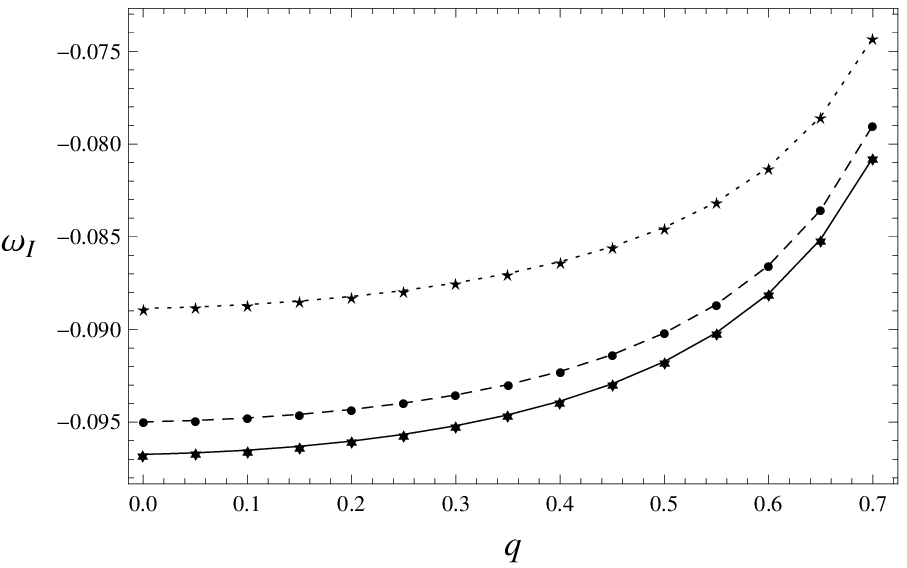}
\includegraphics[width=0.32\linewidth]{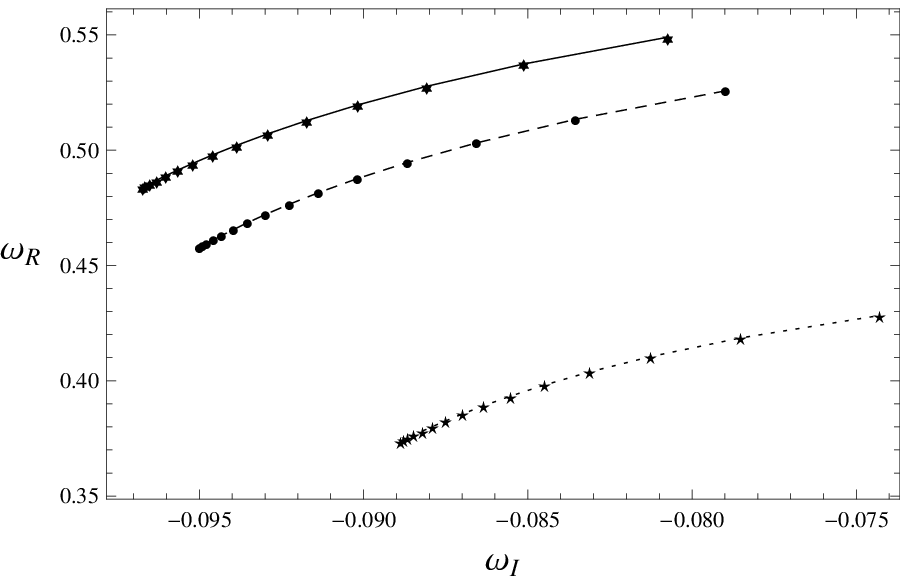}
\end{center}
\caption{\label{wkbwqbardeen} Dependence of the real (left) $\omega_R$ and imaginary (right) $\omega_I$ part of the QNM frequencies of the scalar (solid $\ast$), electromagnetic (dashed $\bullet$) and gravitational (dotted $\star$) perturbative fields on the magnetic charge $q$ of the Bardeen BHs is given for the $l=2$, $n=0$ mode.}
\end{figure*}

\begin{figure*}[t!.]
\begin{center}
\includegraphics[width=0.32\linewidth]{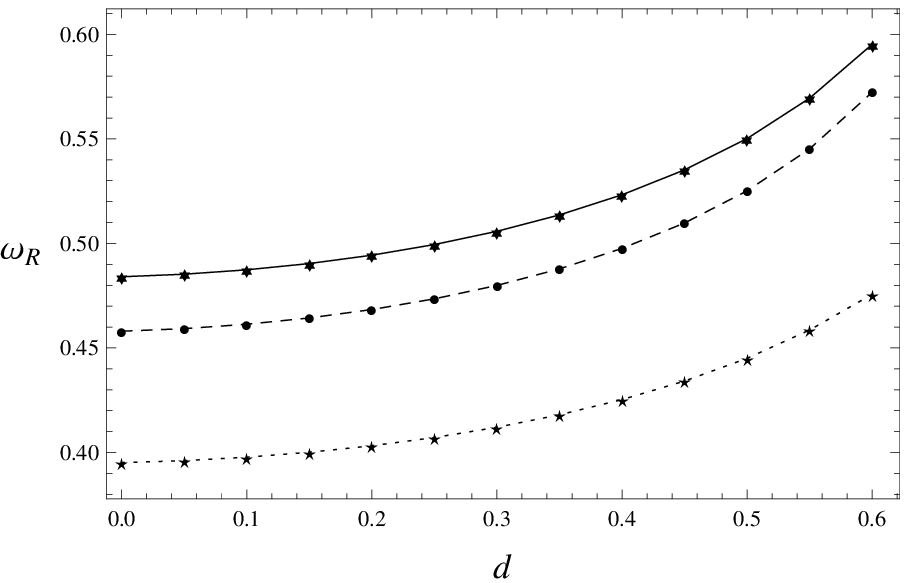}
\includegraphics[width=0.32\linewidth]{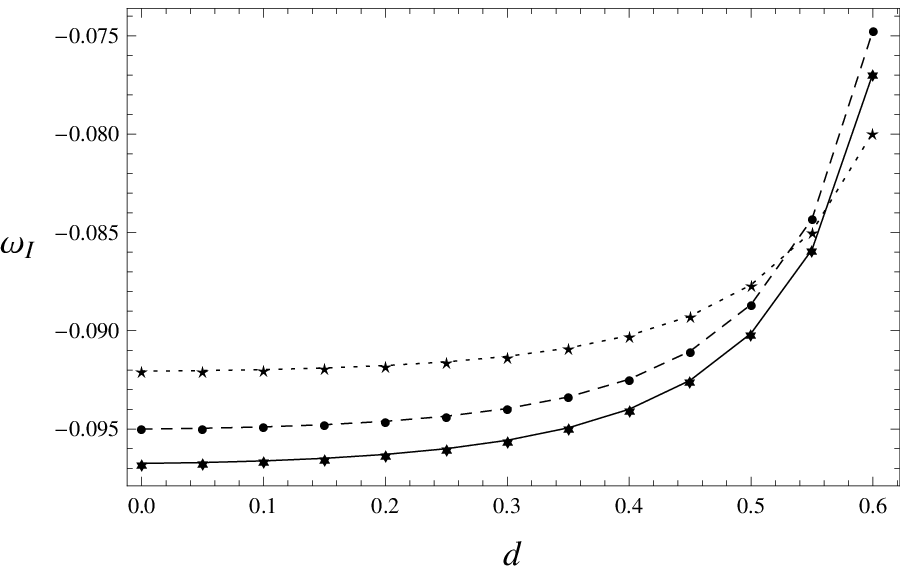}
\includegraphics[width=0.32\linewidth]{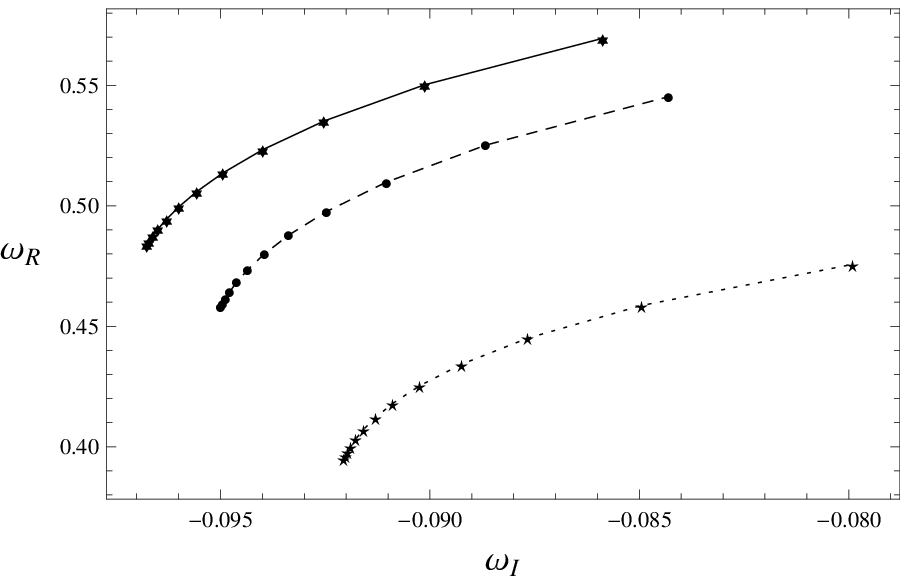}
\end{center}
\caption{\label{wkbwqabg} Dependence of the real (left) $\omega_R$ and imaginary (right) $\omega_I$ part of the QNM frequencies of the scalar (solid $\ast$), electromagnetic (dashed $\bullet$) and gravitational (dotted $\star$) perturbative fields on charge $d$ of the ABG BHs is given for the $l=2$, $n=0$ mode.}
\end{figure*}

In Figs~\ref{wkbwq} -~\ref{wkbwqabg} dependence of the real and imaginary parts of the QNM frequencies are given for the Hayward, Bardeen and ABG BHs in dependence on the spacetime parameter $q_i$ for fixed multipole and overtone $l=2$, $n=0$ parameters.

\begin{table*}[h]
\begin{ruledtabular}
\begin{tabular}{|p{1.5cm}| p{2.4cm}| p{2.4cm}| p{2.4cm}| p{2.4cm} |p{2.4cm}| p{2.4cm}| }
s=0 & \multicolumn{2}{c|}{ Extremal Hayward ($g_c=1.0583$)} & \multicolumn{2}{c|}{Extremal Bardeen ($q_c=0.7698$)} & \multicolumn{2}{c|}{ Extremal ABG ($d_c=0.6342$)}\\
\hline
overtone, multipole & 3rd order WKB & 6th order WKB & 3rd order WKB &  6th order WKB & 3rd order WKB &  6th order WKB \\ \hline
n=0; l=0 & 0.0894 - 0.0978$i$ & 0.1151 - 0.0768$i$ & 0.1045 - 0.0935$i$  & 0.1211 - 0.0776$i$ & 0.1110 - 0.0958$i$ & 0.1271 - 0.0806$i$\\ \hline
n=0; l=1 & 0.3037 - 0.0816$i$ & 0.3072 - 0.0828$i$ & 0.3320 - 0.0789$i$  & 0.3343 - 0.0792$i$ & 0.3513 - 0.0812$i$ & 0.3535 - 0.0815$i$\\ \hline
n=1; l=1 & 0.2609 - 0.2595$i$ & 0.2716 - 0.2583$i$ & 0.2980 - 0.2468$i$  & 0.3046 - 0.2446$i$ & 0.3181 - 0.2532$i$ & 0.3242 - 0.2511$i$\\ \hline
n=1; l=2 & 0.4837 - 0.2486$i$ & 0.4872 - 0.2485$i$ & 0.5337 - 0.2378$i$  & 0.5358 - 0.2375$i$ & 0.5662 - 0.2445$i$ & 0.5680 - 0.2442$i$ \\ \hline
n=1; l=3 & 0.6950 - 0.2463$i$ & 0.6965 - 0.2463$i$ & 0.7607 - 0.2357$i$  & 0.7615 - 0.2356$i$ & 0.8059 - 0.2424$i$ & 0.8066 - 0.2423$i$\\ \hline
n=2; l=2 & 0.4391 - 0.4250$i$ & 0.4432 - 0.4259$i$ & 0.4975 - 0.4042$i$  & 0.4995 - 0.4043$i$ & 0.5307 - 0.4151$i$ & 0.5322 - 0.4153$i$\\ \hline
n=2; l=3 & 0.6623 - 0.4159$i$ & 0.6642 - 0.4160$i$ & 0.7342 - 0.3968$i$  & 0.7351 - 0.3968$i$ & 0.7798 - 0.4080$i$ & 0.7805 - 0.4080$i$ \\ \hline
n=2; l=4 & 0.8769 - 0.4124$i$ & 0.8779 - 0.4124$i$ & 0.9641 - 0.3939$i$  & 0.9646 - 0.3938$i$ & 1.0224 - 0.4050$i$ & 1.0228 - 0.4050$i$ \\
\end{tabular}
\caption{\label{tab_scalar} Extremal Hayward, Bardeen and ABG BH QNM frequencies given for the scalar perturbative field ($s=0$).}
\end{ruledtabular}
\end{table*}
\begin{table*}[h]
\begin{ruledtabular}
\begin{tabular}{|p{1.5cm}| p{2.4cm}| p{2.4cm}| p{2.4cm}| p{2.4cm} |p{2.4cm}| p{2.4cm}| }
s=1 & \multicolumn{2}{c|}{ Extremal Hayward ($g_c=1.0583$)} & \multicolumn{2}{c|}{Extremal Bardeen ($q_c=0.7698$)} & \multicolumn{2}{c|}{ Extremal ABG ($d_c=0.6342$)}\\
\hline
overtone, multipole & 3rd order WKB & 6th order WKB & 3rd order WKB &  6th order WKB & 3rd order WKB &  6th order WKB \\ \hline
n=0; l=1 & 0.2630 - 0.0748$i$ & 0.2663 - 0.0784$i$ & 0.2929 - 0.0733$i$ & 0.2957 - 0.0745$i$ & 0.3114 - 0.0758$i$ & 0.3140 - 0.0767$i$ \\ \hline
n=0; l=2 & 0.4851 - 0.0792$i$ & 0.4861 - 0.0795$i$ & 0.5308 - 0.0763$i$ & 0.5314 - 0.0764$i$ & 0.5623 - 0.0787$i$ & 0.5629 - 0.07876$i$ \\ \hline
n=1; l=1 & 0.2138 - 0.2421$i$ & 0.2277 - 0.2504$i$ & 0.2546 - 0.2316$i$ & 0.2639 - 0.2318$i$ & 0.2740 - 0.2384$i$ & 0.2830 - 0.2377$i$ \\ \hline
n=1; l=2 & 0.4593 - 0.2417$i$ & 0.4631 - 0.2419$i$ & 0.5103 - 0.2320$i$ & 0.5126 - 0.2318$i$ & 0.5423 - 0.2389$i$ & 0.5443 - 0.2387$i$ \\ \hline
n=1; l=3 & 0.6780 - 0.2428$i$ & 0.6795 - 0.2428$i$ & 0.7442 - 0.2327$i$ & 0.7451 - 0.2326$i$ & 0.7890 - 0.2395$i$ & 0.7898 - 0.2395$i$ \\ \hline
n=2; l=2 & 0.4134 - 0.4140$i$ & 0.4183 - 0.4153$i$ & 0.4732 - 0.3947$i$ & 0.4757 - 0.3949$i$ & 0.5058 - 0.4059$i$ & 0.5077 - 0.4062$i$ \\ \hline
n=2; l=3 & 0.6449 - 0.4101$i$ & 0.6469 - 0.4102$i$ & 0.7175 - 0.3919$i$ & 0.7185 - 0.3919$i$ & 0.7627 - 0.4032$i$ & 0.7635 - 0.4032$i$ \\ \hline
n=2; l=4 & 0.8635 - 0.4088$i$ & 0.8646 - 0.4088$i$ & 0.9513 - 0.3909$i$ & 0.9517 - 0.3908$i$ & 1.0092 - 0.4021$i$ & 1.0096 - 0.4021$i$\\
\end{tabular}
\caption{\label{tab_electromagnetic} Extremal Hayward, Bardeen and ABG BH QNM frequencies are given for the electromagnetic perturbative fields ($s=1$).}
\end{ruledtabular}
\end{table*}

\begin{table*}[h]
\begin{ruledtabular}
\begin{tabular}{|p{1.5cm}| p{2.4cm}| p{2.4cm}| p{2.4cm}| p{2.4cm} |p{2.4cm}| p{2.4cm}| }
s=2 & \multicolumn{2}{c|}{ Extremal Hayward ($g_c=1.0583$)} & \multicolumn{2}{c|}{Extremal Bardeen ($q_c=0.7698$)} & \multicolumn{2}{c|}{ Extremal ABG ($d_c=0.6342$)}\\
\hline
overtone, multipole & 3rd order WKB & 6th order WKB & 3rd order WKB &  6th order WKB & 3rd order WKB &  6th order WKB \\ \hline
n=0; l=2 & 0.3944 - 0.0735$i$ & 0.3954 - 0.0753$i$  & 0.4321 - 0.0714$i$ & 0.4328 - 0.0719$i$ & 0.4689 - 0.0818$i$ & 0.4698 - 0.0819$i$\\ \hline
n=0; l=3 & 0.6342 - 0.0779$i$ & 0.6347 - 0.0780$i$  & 0.6916 - 0.0750$i$ & 0.6919 - 0.0751$i$ & 0.7374 - 0.0803$i$ & 0.7377 - 0.0803$i$\\ \hline
n=1; l=2 & 0.3620 - 0.2256$i$ & 0.3675 - 0.2349$i$  & 0.4083 - 0.2180$i$ & 0.4100 - 0.2209$i$ & 0.4445 - 0.2505$i$ & 0.4473 - 0.2495$i$\\ \hline
n=1; l=3 & 0.6156 - 0.2359$i$ & 0.6174 - 0.2360$i$  & 0.6767 - 0.2267$i$ & 0.6778 - 0.2266$i$ & 0.7222 - 0.2425$i$ & 0.7232 - 0.2424$i$\\ \hline
n=1; l=4 & 0.8422 - 0.2394$i$ & 0.8430 - 0.2394$i$  & 0.9210 - 0.2297$i$ & 0.9214 - 0.2296$i$ & 0.9788 - 0.2412$i$ & 0.9792 - 0.2412$i$\\ \hline
n=2; l=2 & 0.3037 - 0.3901$i$ & 0.3218 - 0.4190$i$  & 0.3657 - 0.3736$i$ & 0.3656 - 0.3888$i$ & 0.4030 - 0.4290$i$ & 0.4043 - 0.4294$i$\\ \hline
n=2; l=3 & 0.5803 - 0.3992$i$ & 0.5829 - 0.3997$i$  & 0.6484 - 0.3823$i$ & 0.6497 - 0.3823$i$ & 0.6935 - 0.4090$i$ & 0.6945 - 0.4091$i$\\ \hline
n=2; l=4 & 0.8155 - 0.4025$i$ & 0.8167 - 0.4025$i$  & 0.8994 - 0.3854$i$ & 0.9000 - 0.3854$i$ & 0.9572 - 0.4047$i$ & 0.9577 - 0.4047$i$\\
\end{tabular}
\caption{\label{tab_graviton} Extremal Hayward, Bardeen and ABG BH QNM frequencies are given for the gravitational perturbative fields ($s=2$).}
\end{ruledtabular}
\end{table*}

It is well known that the $l=0$ modes admit only $n=0$ overtones. One can see from the Tab.~\ref{wkbscalar} that for this mode the third and the sixth order WKB results demonstrate large relative error. For the $l=0$, $n=0$ mode, such an error between the 3th and 6th order WKB approximations has been observed also for the Schwarzschild BHs~\cite{Konoplya2}.


\begin{figure*}[t!.]
\begin{center}
\includegraphics[width=0.32\linewidth]{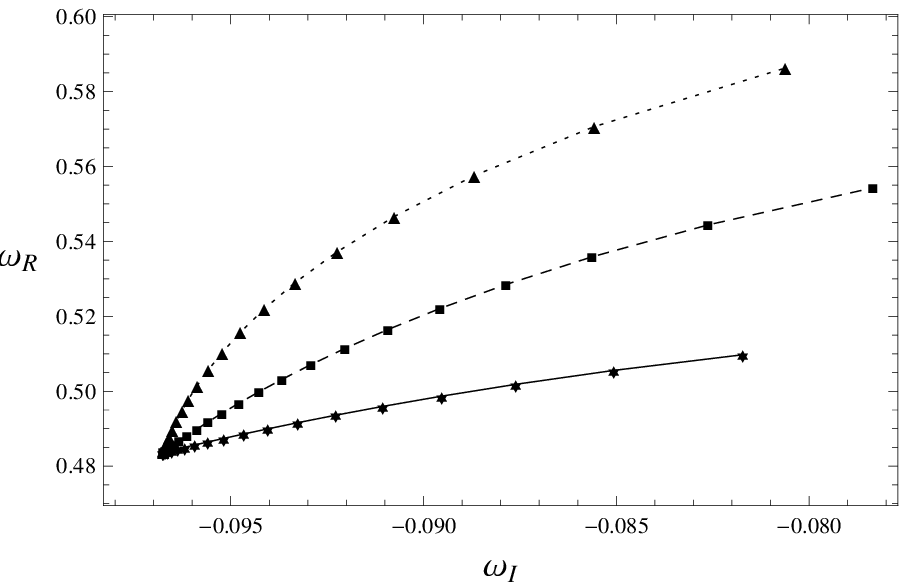}
\includegraphics[width=0.32\linewidth]{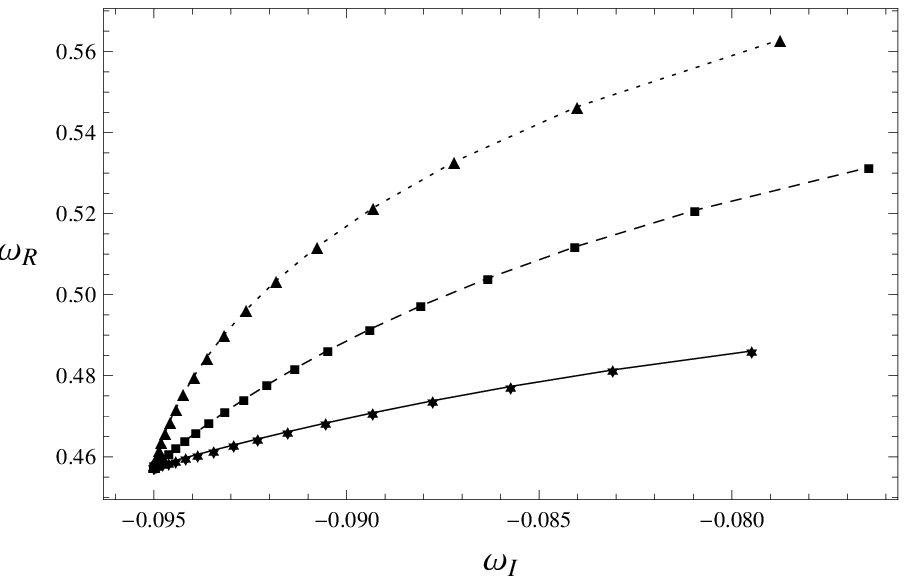}
\includegraphics[width=0.32\linewidth]{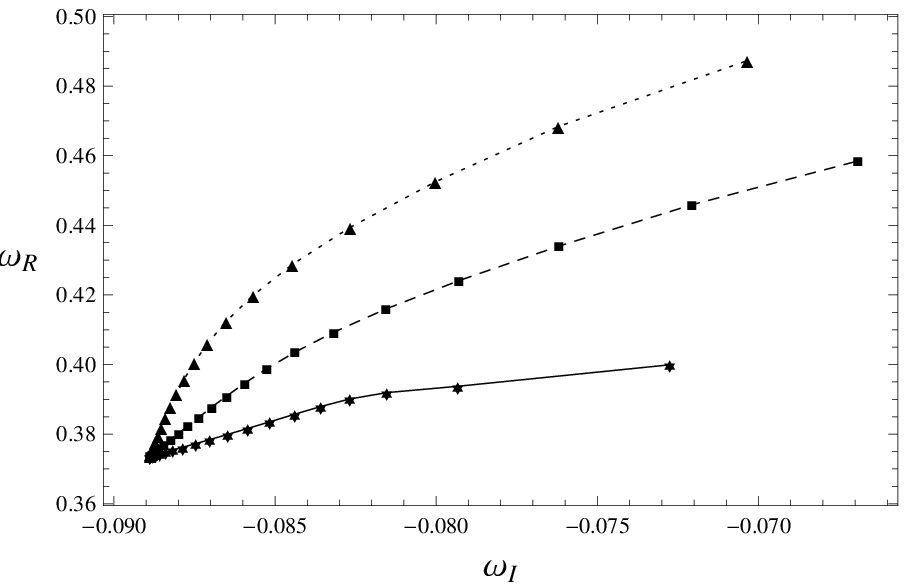}
\includegraphics[width=0.32\linewidth]{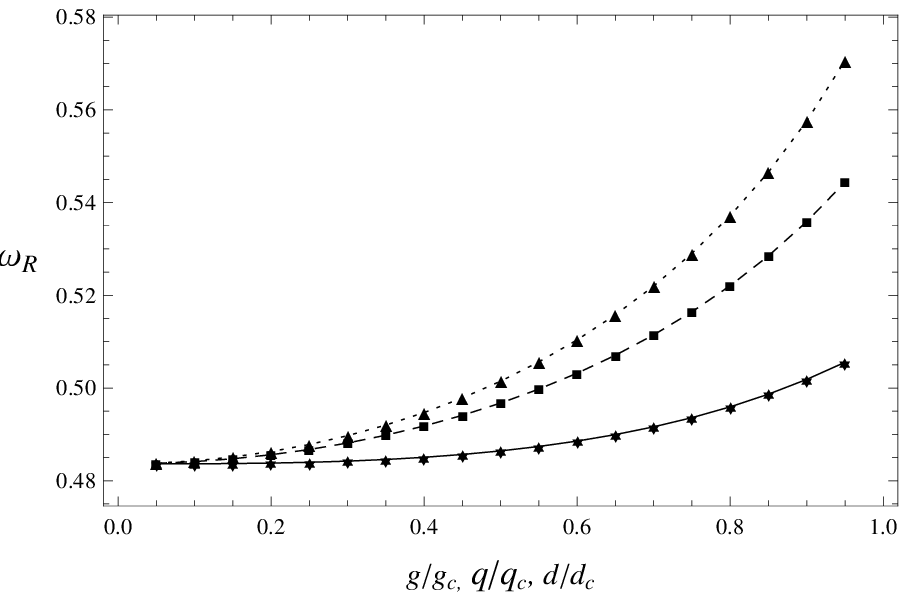}
\includegraphics[width=0.32\linewidth]{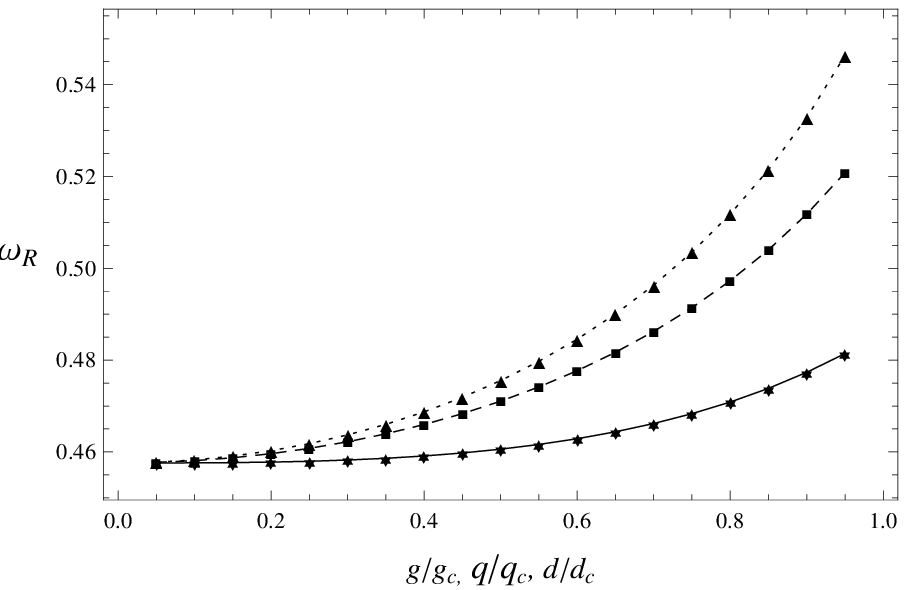}
\includegraphics[width=0.32\linewidth]{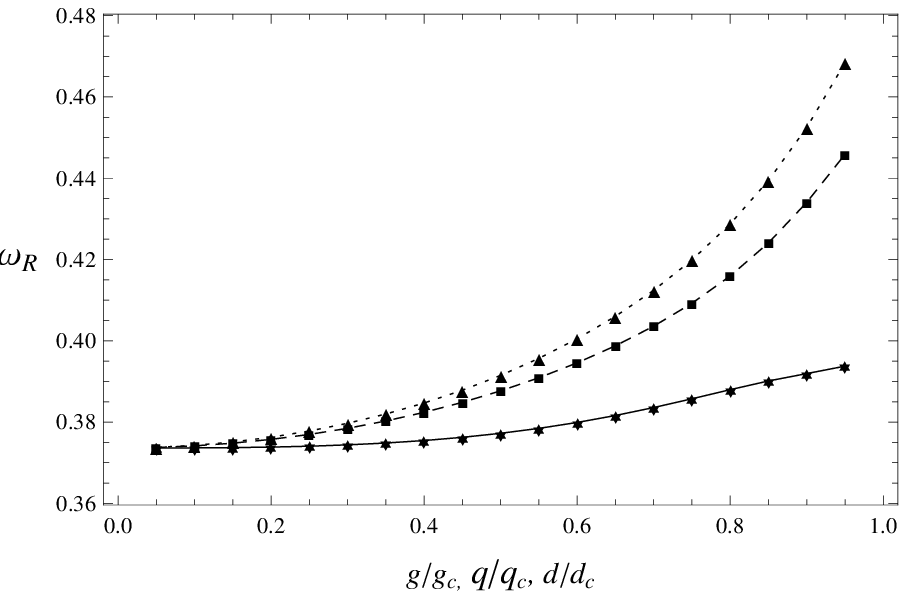}
\includegraphics[width=0.32\linewidth]{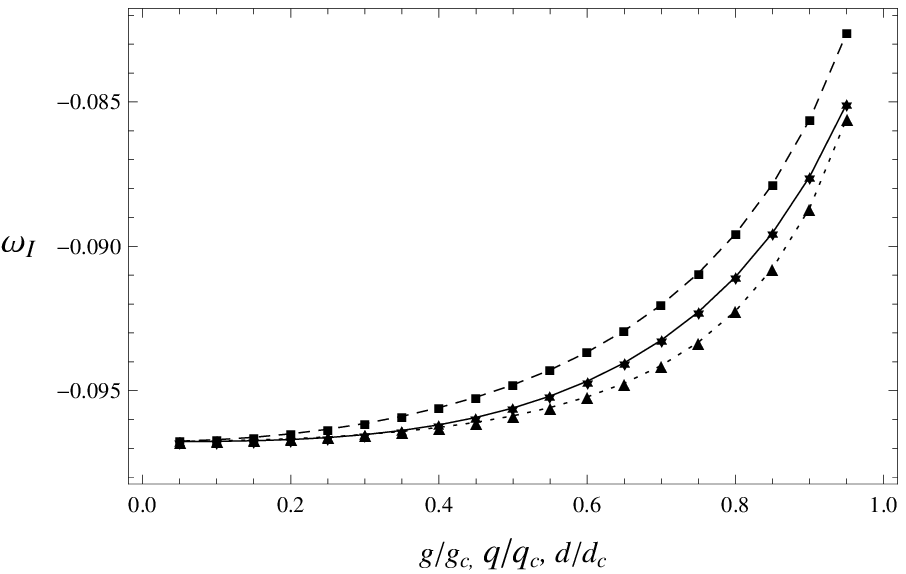}
\includegraphics[width=0.32\linewidth]{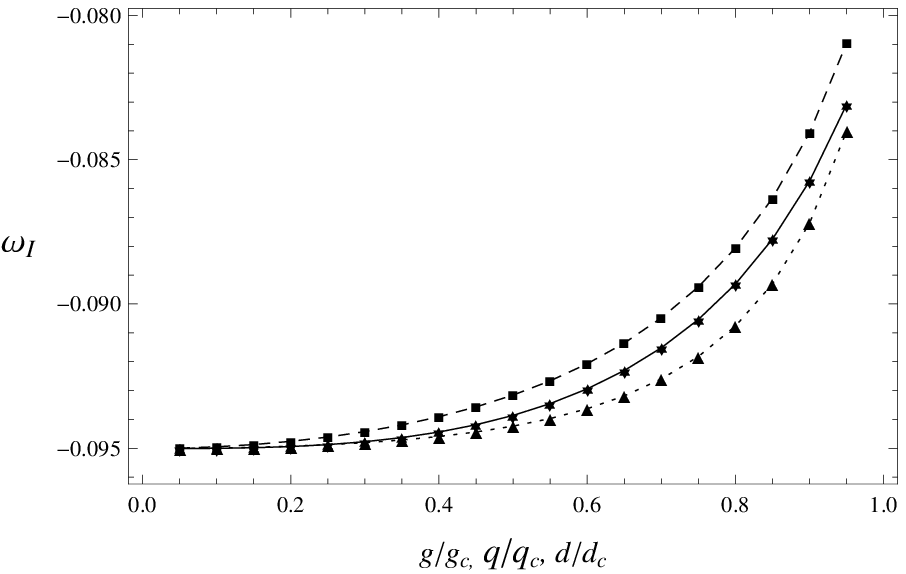}
\includegraphics[width=0.32\linewidth]{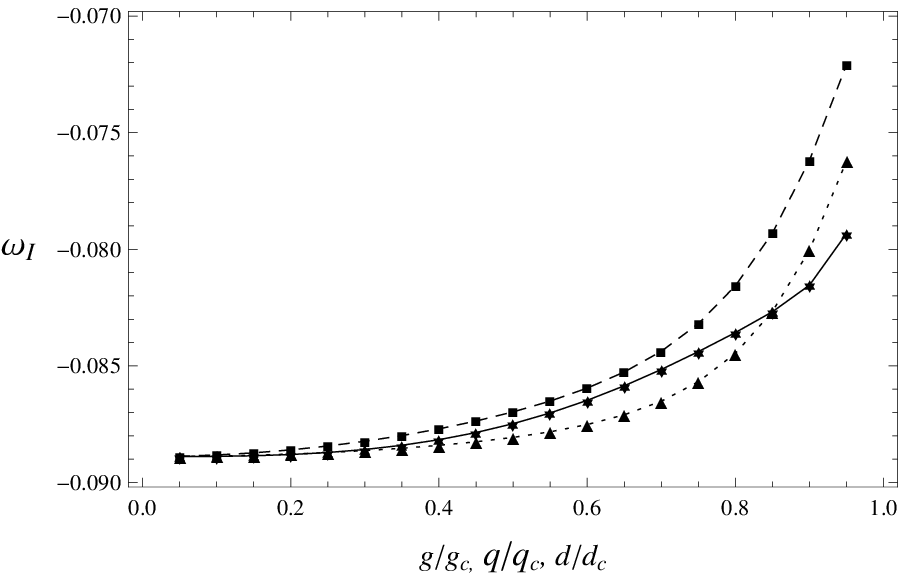}
\end{center}
\caption{\label{dep} Plots showing the behavior of $l=2$, $n=0$ QNM frequencies in the complex $\omega$ plane for the scalar, electromagnetic and gravitational perturbative fields (from left to right) in the Hayward (solid $\ast$), Bardeen (dashed $\blacksquare$) and ABG (dotted $\blacktriangle$) BH spacetimes.}
\end{figure*}
In Fig.~\ref{dep} intersection of all the curves corresponds to the QNM frequency related to the Schwarzschild BHs ($g=0$, $q=0$ and $d=0$).

\begin{table*}[h]
\begin{ruledtabular}
\begin{tabular}{|c| p{2.4cm}| p{2.4cm}| p{2.4cm}| p{2.4cm} |p{2.4cm}| p{2.4cm}| }
$s=0$& \multicolumn{2}{c|}{ Hayward } & \multicolumn{2}{c|}{ Bardeen } & \multicolumn{2}{c|}{ ABG }\\
\hline
$g/g_c$, $q/q_c$, $d/d_c$ & 3rd order WKB & 6th order WKB & 3rd order WKB &  6th order WKB & 3rd order WKB &  6th order WKB \\ \hline
0.1 & 0.2911 - 0.0980$i$ & 0.2929 - 0.0977$i$ & 0.2914 - 0.0979$i$  & 0.2932 - 0.0977$i$ & 0.2915 - 0.0980$i$ & 0.2933 - 0.0977$i$\\ \hline
0.2 & 0.2912 - 0.0979$i$ & 0.2930 - 0.0977$i$ & 0.2924 - 0.0977$i$  & 0.2941 - 0.0975$i$ & 0.2928 - 0.0979$i$ & 0.2946 - 0.0976$i$\\ \hline
0.3 & 0.2915 - 0.0977$i$ & 0.2933 - 0.0975$i$ & 0.2939 - 0.0973$i$  & 0.2957 - 0.0971$i$ & 0.2949 - 0.0977$i$ & 0.2967 - 0.0975$i$\\ \hline
0.4 & 0.2919 - 0.0973$i$ & 0.2938 - 0.0972$i$ & 0.2962 - 0.0966$i$  & 0.2980 - 0.0966$i$ & 0.2980 - 0.0974$i$ & 0.2998 - 0.0972$i$\\ \hline
0.5 & 0.2927 - 0.0966$i$ & 0.2947 - 0.0966$i$ & 0.2993 - 0.0957$i$  & 0.3011 - 0.0957$i$ & 0.3022 - 0.0969$i$ & 0.3039 - 0.0968$i$\\ \hline
0.6 & 0.2939 - 0.0955$i$ & 0.2960 - 0.0957$i$ & 0.3033 - 0.0945$i$  & 0.3051 - 0.0946$i$ & 0.3077 - 0.0961$i$ & 0.3094 - 0.0961$i$\\ \hline
0.7 & 0.2956 - 0.0938$i$ & 0.2979 - 0.0943$i$ & 0.3083 - 0.0927$i$  & 0.3102 - 0.0929$i$ & 0.3148 - 0.0949$i$ & 0.3165 - 0.0950$i$\\ \hline
0.8 & 0.2978 - 0.0913$i$ & 0.3005 - 0.0920$i$ & 0.3146 - 0.0900$i$  & 0.3166 - 0.0903$i$ & 0.3239 - 0.0928$i$ & 0.3257 - 0.0930$i$\\ \hline
0.9 & 0.3006 - 0.0875$i$ & 0.3038 - 0.0885$i$ & 0.3225 - 0.0859$i$  & 0.3247 - 0.0862$i$ & 0.3358 - 0.0890$i$ & 0.3378 - 0.0893$i$\\ \hline
0.99 & 0.3034 - 0.0823$i$ & 0.3069 - 0.0835$i$ & 0.3310 - 0.0798$i$  & 0.3333 - 0.0801$i$ & 0.3496 - 0.0823$i$ & 0.3518 - 0.0826$i$\\ \hline
1.0 & 0.3037 - 0.0816$i$ & 0.3072 - 0.0828$i$ & 0.3320 - 0.0789$i$  & 0.3343 - 0.0792$i$ & 0.3513 - 0.0812$i$ & 0.3535 - 0.0815$i$\\
\end{tabular}
\caption{\label{tab_scalar_dep_q} Dependence of the Hayward, Bardeen and ABG BH QNM frequencies on the parameters $q_i/q_c$ are given for the scalar perturbative fields ($s=0$).}
\end{ruledtabular}
\end{table*}

\section{Scattering and greybody factor}

The Schr\"{o}dinger-like equation (\ref{16}) can be applied to scattering of waves in the regular BH spacetimes by using the standard techniques for tunneling in quantum mechanics. The asymptotic solutions of the Schr\"{o}dinger equation (\ref{16}) read:
\begin{eqnarray}\label{s1}
\psi=A(\omega)e^{-i\omega r_\ast}+B(\omega)e^{i\omega r_\ast}, \quad & r_\ast\rightarrow-\infty,\nonumber\\
\psi=C(\omega)e^{-i\omega r_\ast}+D(\omega)e^{i\omega r_\ast}, \quad & r_\ast\rightarrow+\infty.
\end{eqnarray}
For waves incoming towards the regular BHs from infinity there is necessarily $B(\omega)=0$. The reflection
amplitude is given by $R(\omega)=D(\omega)/C(\omega)$, while the transmission amplitude reads  $T(\omega)=A(\omega)/C(\omega)$. Therefore, we can write
\begin{eqnarray}\label{s2}
\psi=T(\omega)e^{-i\omega r_{\ast}}, \quad & r_{\ast}\rightarrow-\infty\ ,\nonumber\\
\psi=e^{-i\omega r_\ast}+R(\omega)e^{i\omega r_\ast}, \quad & r_\ast\rightarrow+\infty\ .
\end{eqnarray}
The square of the amplitude of the wave function at a particular point of spacetime determines probability of finding it in the given point. The wave incoming towards a regular BH is partially transmitted and partially reflected by the potential barrier. Probability of finding this wave in the whole region above the horizon is always equal to one, therefore the condition
\begin{eqnarray}\label{s3}
|R(\omega)|^2+|T(\omega)|^2=1\ ,
\end{eqnarray}
has to be satisfied. Here we should consider following three cases:\\
(i) $\omega^2$ is much less then the maximum of the effective potential $V(r_0)$ ($\omega^2\ll V(r_0)$); \\
(ii) $\omega^2$ is of the same order as the maximum of the effective potential $V(r_0)$ ($\omega^2\approx V(r_0)$); \\
(iii) $\omega^2$ is much larger then the maximum of the effective potential $V(r_0)$ ($\omega^2\gg V(r_0)$).

\begin{figure*}[h!.]
\begin{center}
\includegraphics[width=0.32\linewidth]{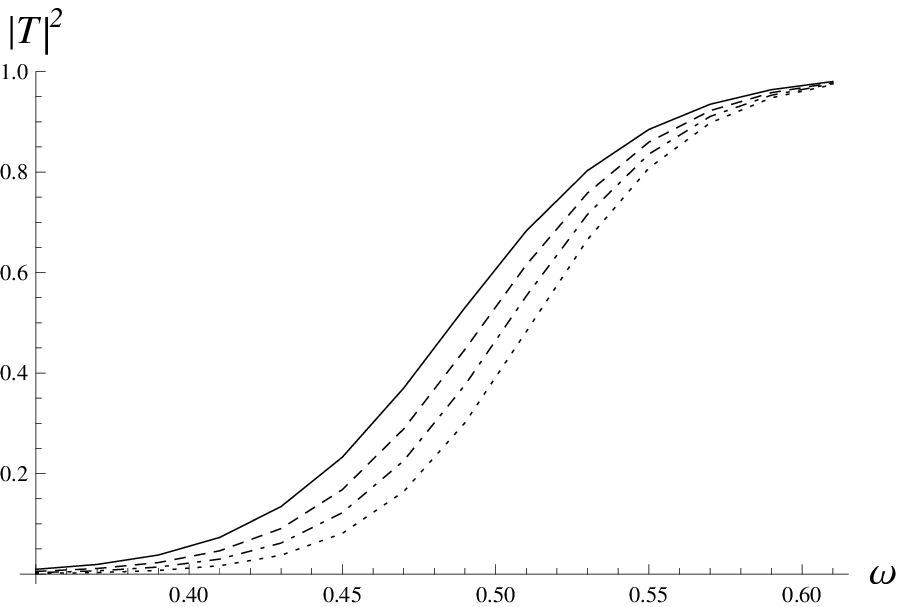}
\includegraphics[width=0.32\linewidth]{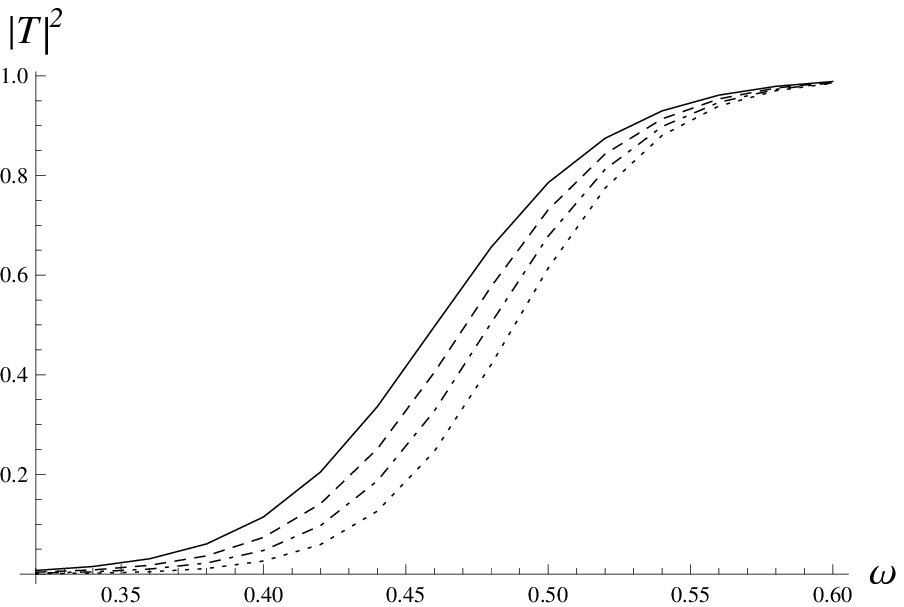}
\includegraphics[width=0.32\linewidth]{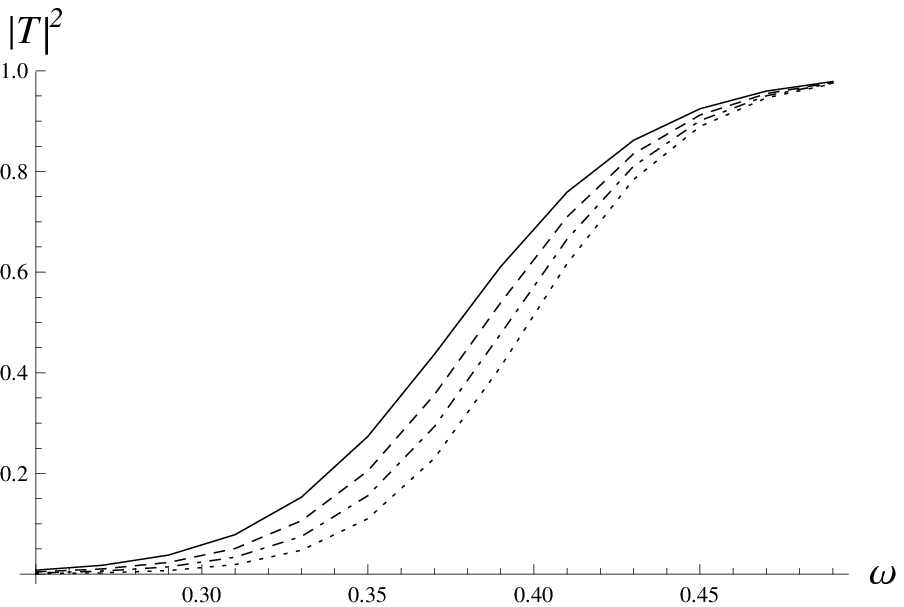}
\includegraphics[width=0.32\linewidth]{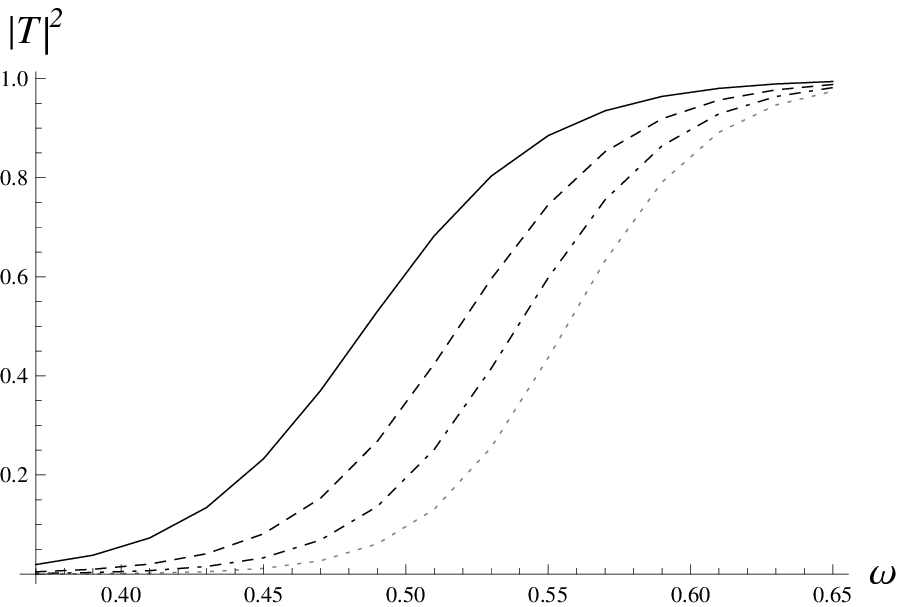}
\includegraphics[width=0.32\linewidth]{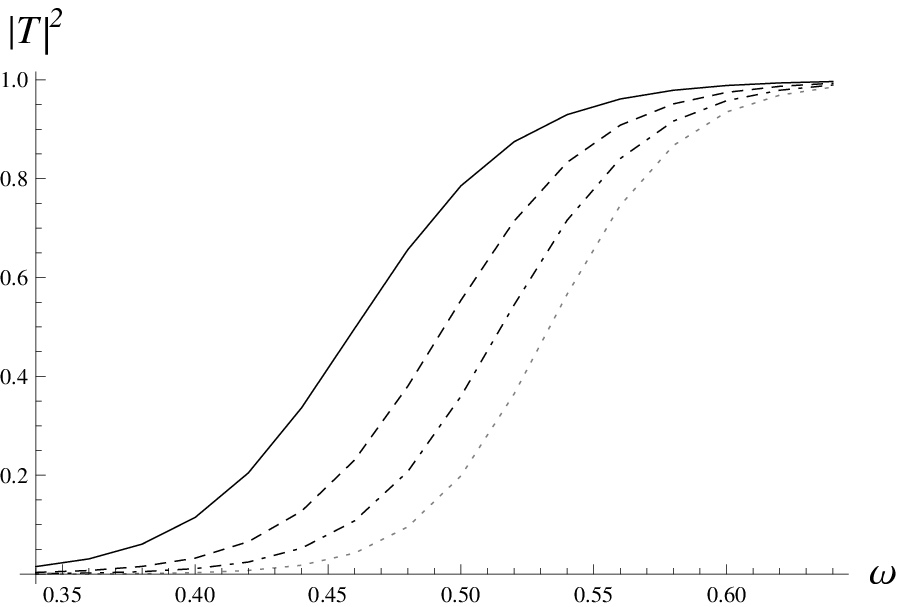}
\includegraphics[width=0.32\linewidth]{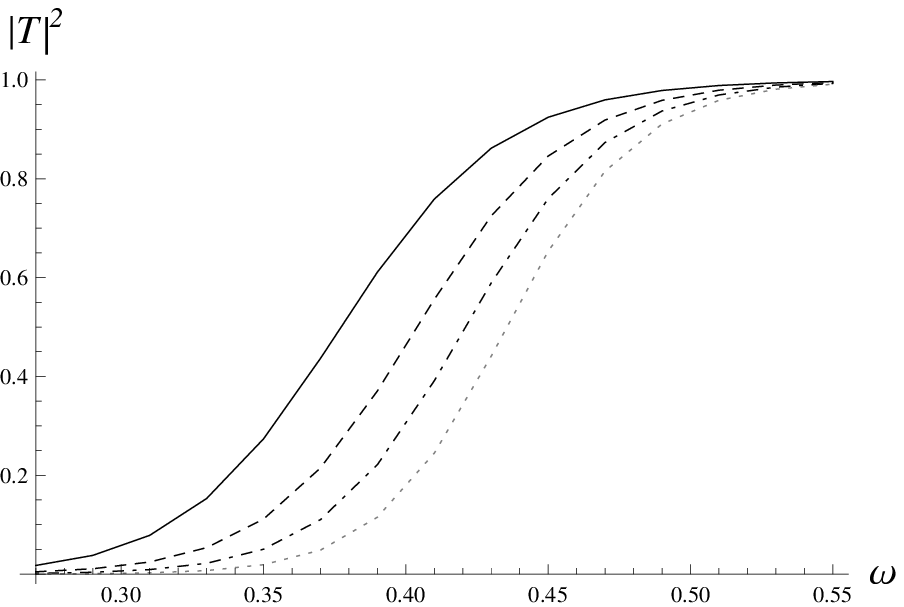}
\includegraphics[width=0.32\linewidth]{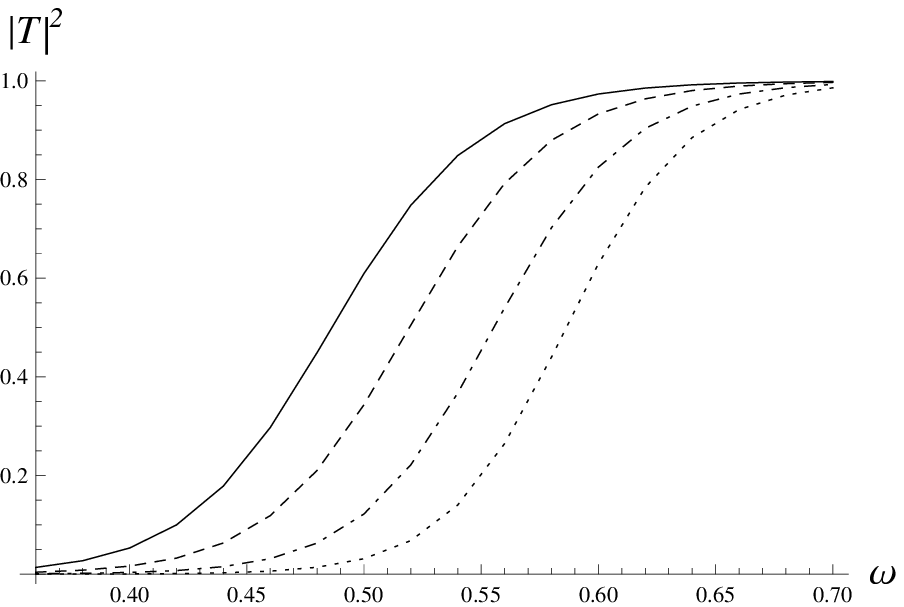}
\includegraphics[width=0.32\linewidth]{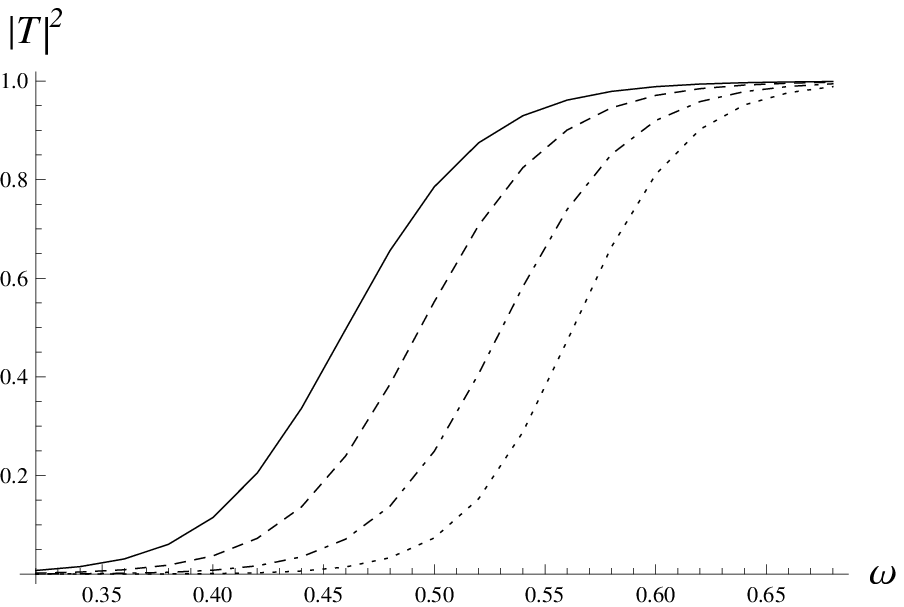}
\includegraphics[width=0.32\linewidth]{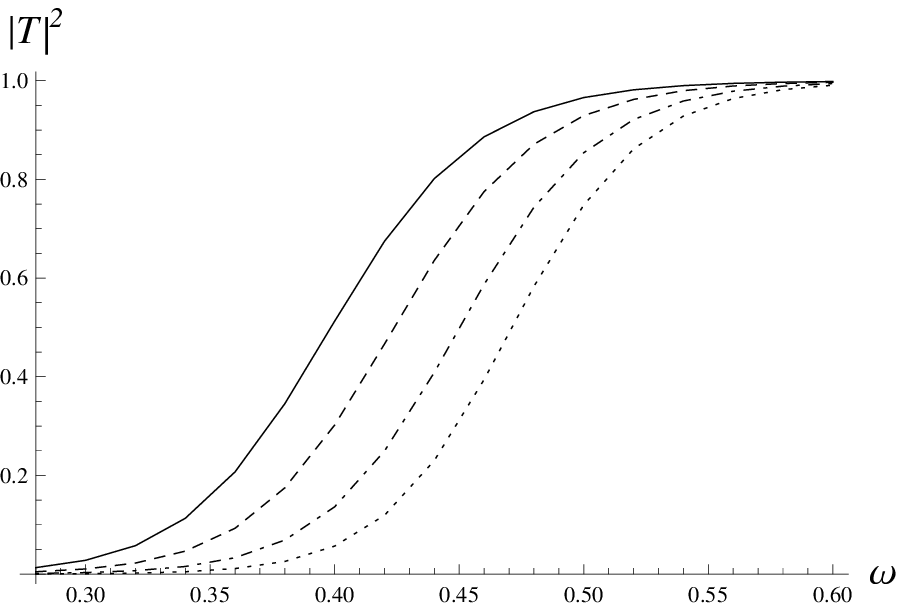}
\end{center}
\caption{\label{transmission} Transmission coefficients for the scalar, electromagnetic and gravitational (from left to right) fields in the  Hayward, Bardeen and ABG BH spacetimes (from top to bottom). There is $\ell=2$, $q=0$ (solid), $q=0.75q_c$ (dashed), $q=0.9q_c$ (dotdashed), $q=q_c$ (dotted).}
\end{figure*}

\begin{figure*}[h!.]
\begin{center}
\includegraphics[width=0.32\linewidth]{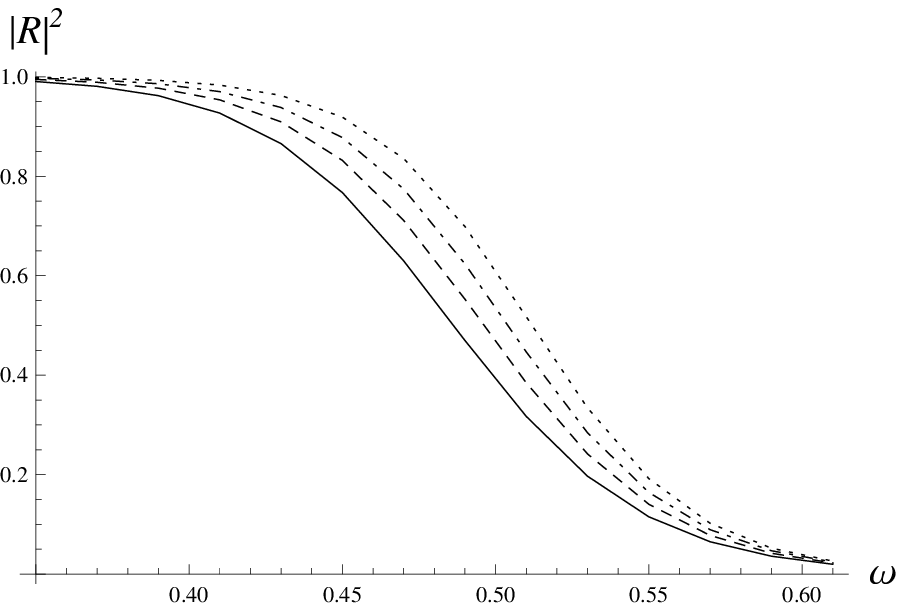}
\includegraphics[width=0.32\linewidth]{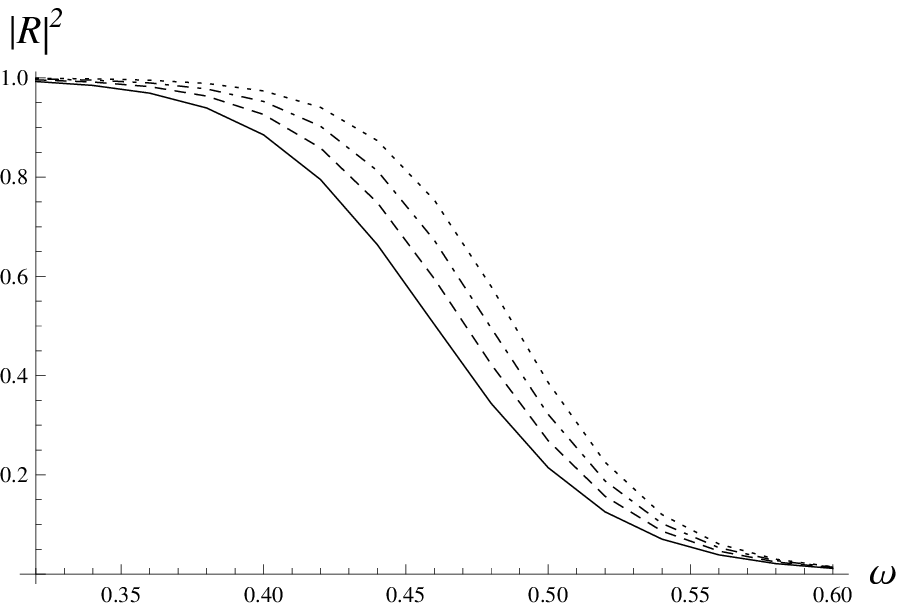}
\includegraphics[width=0.32\linewidth]{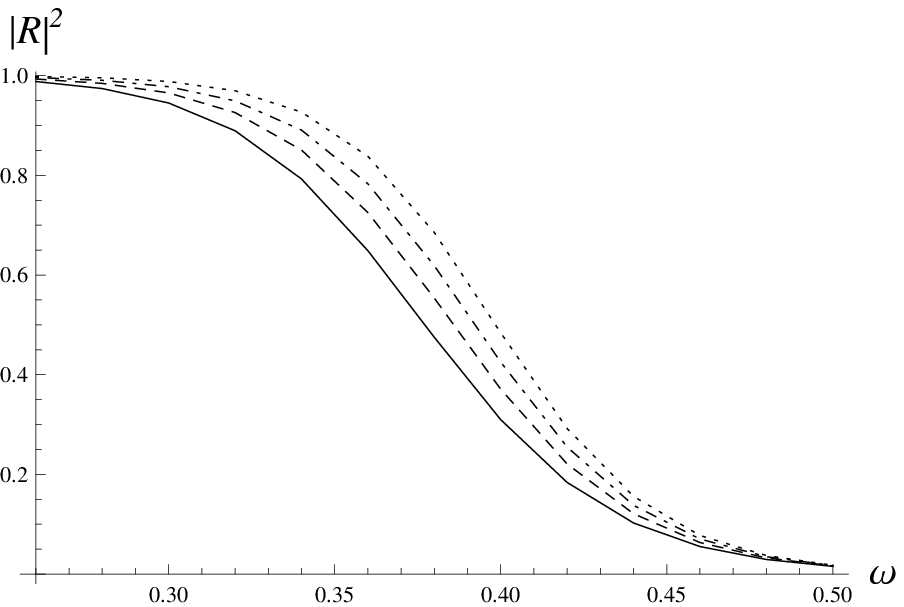}
\includegraphics[width=0.32\linewidth]{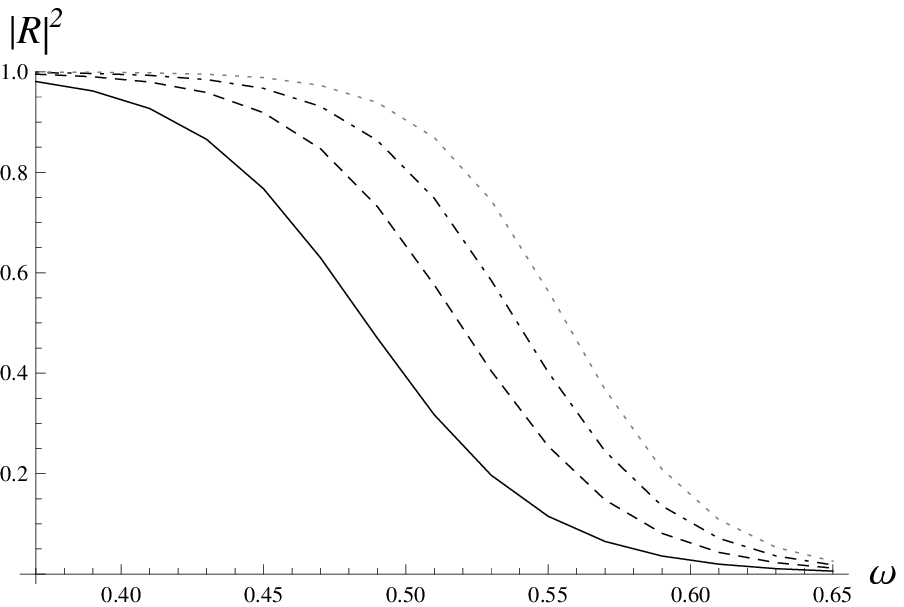}
\includegraphics[width=0.32\linewidth]{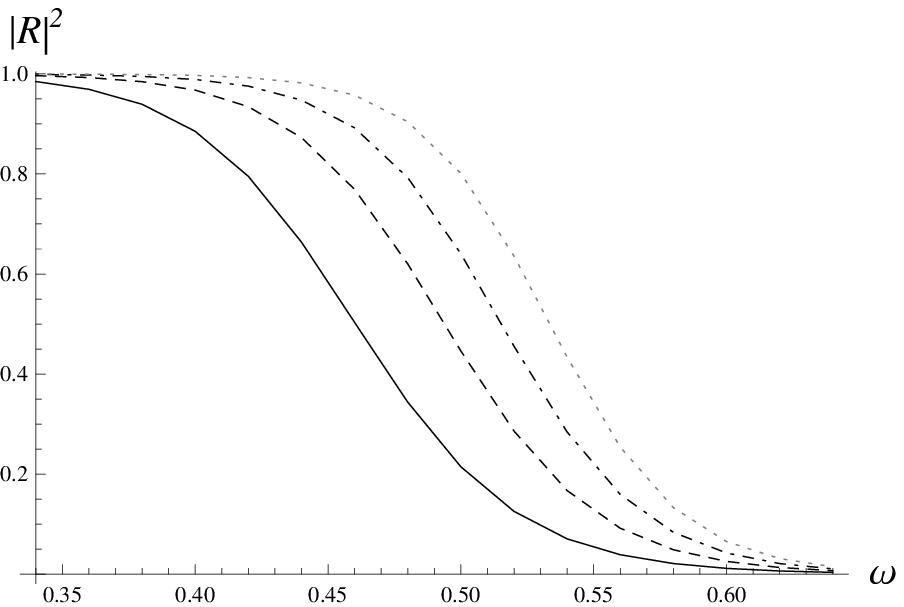}
\includegraphics[width=0.32\linewidth]{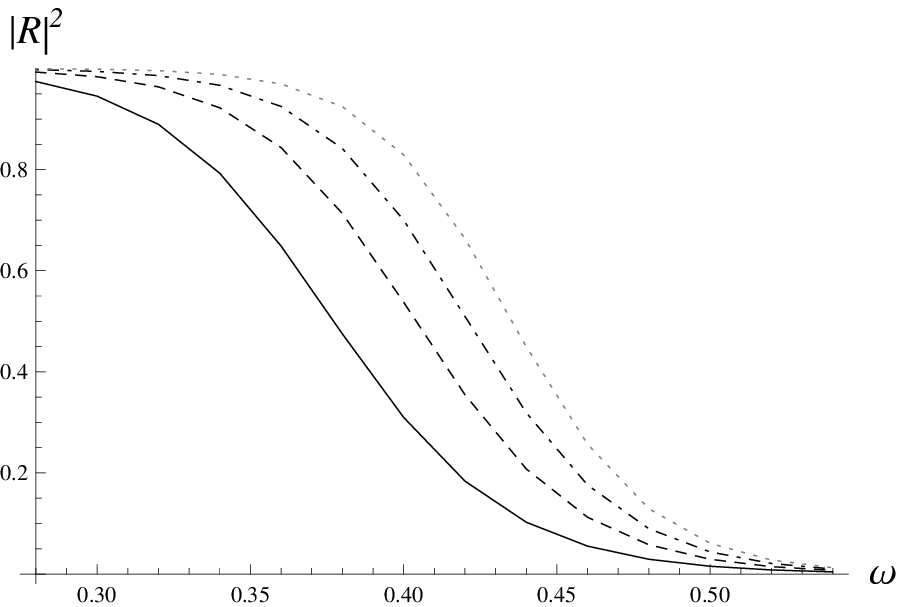}
\includegraphics[width=0.32\linewidth]{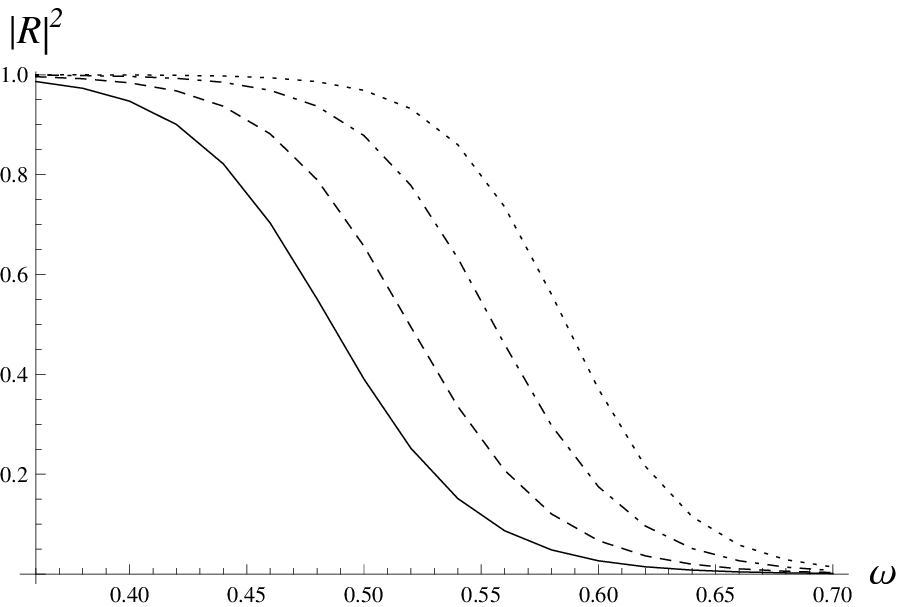}
\includegraphics[width=0.32\linewidth]{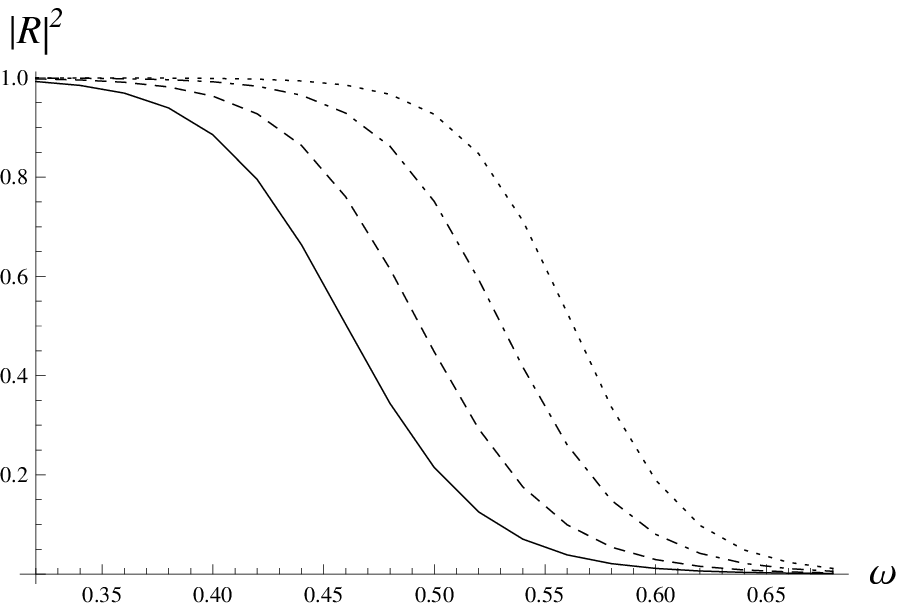}
\includegraphics[width=0.32\linewidth]{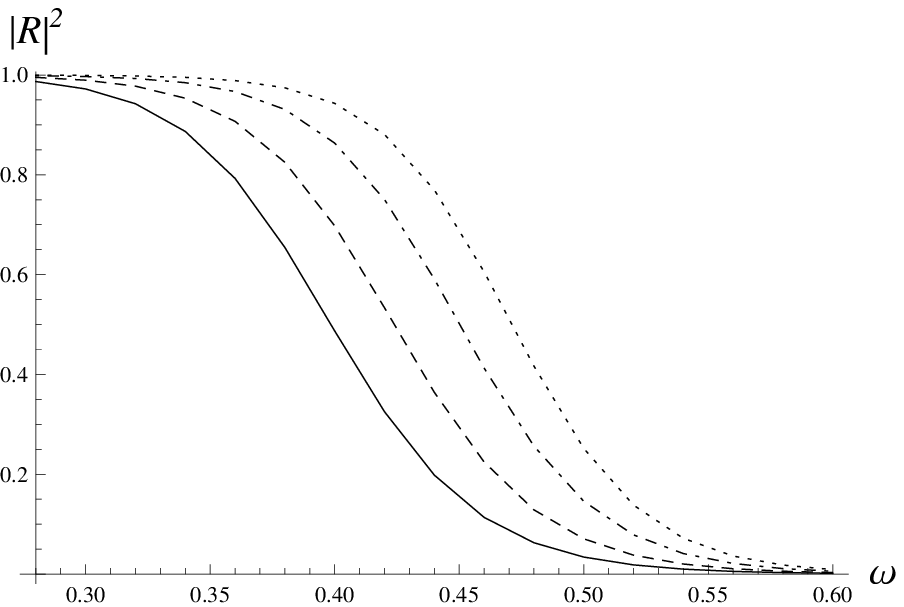}
\end{center}
\caption{\label{reflection} (color online). Reflection coefficients for the scalar, electromagnetic and gravitational (from left to right) fields in the Hayward, Bardeen and ABG BH spacetimes (from top to bottom). There is $\ell=2$, $q=0$ (solid), $q=0.75q_c$ (dashed), $q=0.9q_c$ (dotdashed), $q=q_c$ (dotted).}
\end{figure*}

In the first case, the greybody factor. i.e., the transmission coefficient, is close to zero and reflection coefficient is nearly equal to one. The third case is opposite to the first one, namely the transmission coefficient is close to one, while the reflection coefficient is close to zero~\cite{Zhidenko1}. The most interesting case is the second one, $\omega^2\simeq V(r_0)$. It is known that the WKB method has high accuracy when classical turning points (solution of the equation $\omega^2-V(r_0)=0$) are very close to each other. This occurs namely when the condition $\omega^2\sim V(r_0)$ is satisfied. Therefore, in this case we can find the transmission and reflection coefficients by applying the sixth order WKB approach (see~\cite{Zhidenko1} and references therein). Then the reflection amplitude reads
\begin{eqnarray}\label{s4}
R(\omega)=\left(1+e^{-i2\pi K(\omega)}\right)^{-1/2}\ ,
\end{eqnarray}
where
\begin{eqnarray}\label{s5}
K(\omega)=\frac{i\left(\omega^2-V(r_0)\right)}{\sqrt{-2V''(r_0)}}+\sum_{i=2}^{6}\Lambda_i\ .
\end{eqnarray}
From condition~(\ref{s3}) we can find the expression for the transmission coefficient that reads
\begin{eqnarray}\label{s6}
\left|T(\omega)\right|^2=1-\left|\left(1+e^{-i2\pi K(\omega)}\right)^{-1/2}\right|^2\ .
\end{eqnarray}
In Figs~\ref{transmission} and~\ref{reflection} results of numerical calculations of the transmission and reflection coefficients for the scalar, electromagnetic and gravitational fields with the multipole number $\ell=2$ are shown for the regular Hayward, Bardeen and ABG BH spacetimes.

\section{Conclusions}

We have studied QNMs of the linear "axial" scalar, electromagnetic and gravitational perturbations in the Hayward, Bardeen and ABG regular BH spacetimes by using the sixth order WKB approximation. Calculations have shown that increasing of the spacetime charge parameter implies monotonic increasing of the real part of QNM frequency and monotonic decreasing of the imaginary part of QNM frequency, i.e., the damping rate of the wave decreases. It means that in the regular BHs oscillators are better (slowly damped) than  in the field of Schwarzschild BHs. In the Hayward and Bardeen regular BH spacetimes, damping of the QNMs is always largest for the scalar fields, mediate for the electromagnetic fields and smallest for the gravitational fields. However, for the ABG BH spacetimes the situation is more complex, as the scalar fields have damping smaller than the electromagnetic fields, if the charge parameter of these spacetimes $d > 0.56$.

It has been shown that all of the considered regular BHs the linear "axial" perturbative scalar, electromagnetic and gravitational fields are stable.

The greybody factors of scattering of scalar, electromagnetic gravitational waves, namely the transmission and reflection coefficients through the potential barrier (effective potential) of the regular BH spacetimes have been calculated numerically under the condition $\omega^2\simeq V(r_0)$ related to the WKB approximation, considering $\omega$ is real. In Figs.~\ref{transmission} and~\ref{reflection} it is demonstrated that for the gravitational perturbative fields probability of the wave transmission through the potential barrier is larger than for the scalar and electromagnetic ones. Increase in the value of the spin of the perturbative fields weakens the potential barrier as related to transmission of waves through. Furthermore, an increase in charge of the regular BHs decreases the transmission and increases the reflection of the wave through potential barrier. One can conclude that the regular Hayward and ABG BHs are favoured in terms of transmission and reflection of the wave through their potential barriers, respectively.

\begin{acknowledgments}
The authors would like to express their acknowledgements for the Institutional support of the Faculty of Philosophy and Science of the Silesian University at Opava, the internal student grant of the Silesian University SGS/23/2013. Z.S. acknowledges the Albert Einstein Centre for Gravitation and Astrophysics under the Czech Science Foundation No.~14-37086G. The research of A.A. and B.A. is supported in part by Projects  No.~F2-FA-F113, No.~EF2-FA-0-12477, and No.~F2-FA-F029 of the UzAS, and by the ICTP through Grants No.~OEA-PRJ-29 and No.~OEA-NET-76 and by the Volkswagen Stiftung, Grant No.~866. B.T. would like to thank Alexander Zhidenko for useful discussions and providing the \textit{Mathematica} file for the sixth order WKB approximation.
\end{acknowledgments}


%
%

%
%
%
%

\label{lastpage}


\begin{thebibliography}{99}


%
\bibitem{Regge}{T. Regge and J. A. Wheeler, Phys. Rev. {\bf 108}, 1063 (1957).}
%
\bibitem{Zerilli}{F. J. Zerilli, Phys. Rev. Lett. {\bf 24}, 737 (1970);\\ Phys. Rev. D {\bf 2}, 2141 (1970).}
%
\bibitem{Vishveshwara}{C. V. Vishveshwara, Phys. Rev. D {\bf 1}, 2870 (1970).}
%
\bibitem{Nagar}{A. Nagar and L. Rezzolla, Classical Quantum Gravity {\bf 22}, R167 (2005).}
%
\bibitem{Boonserm}{P. Boonserm and M. Visser, JHEP {\bf 1103}, 073 (2011) [arXiv:1005.4483 [math-ph]].}
%
\bibitem{Kokkotas}{K. D. Kokkotas and B. G. Schmidt, Living Rev. Rel. {\bf 2}, 2 (1999) [arXiv:gr-qc/9909058].}
%
\bibitem{Berti}{E. Berti, V. Cardoso, and A. O. Starinets, Classical Quantum Gravity {\bf 26}, 163001 (2009) [arXiv:0905.2975 [gr-qc]].}
%
\bibitem{Konoplya0}{R. A. Konoplya and A. Zhidenko, Rev. Mod. Phys. {\bf 83}, 793 (2011) [arXiv:1102.4014 [gr-qc]].}
%
\bibitem{Meq1}{H. Kodama and A. Ishibashi, Prog. Theor. Phys. {\bf 110}, 701 (2003) [arXiv:hep-th/0305147].}
%
\bibitem{Meq2}{A. Ishibashi and H. Kodama, Prog. Theor. Phys. {\bf 110}, 901 (2003) [arXiv:hep-th/0305185].}
%
\bibitem{Meq3}{H. Kodama and A. Ishibashi, Prog. Theor. Phys. {\bf 111}, 29 (2004) [arXiv:hep-th/0308128].}
%
\bibitem{Schutz}{B. F. Schutz and C. M. Will, Astrophys. J. {\bf 291}, L33 (1985).}
%
\bibitem{Iyer}{S. Iyer and C. M. Will, Phys. Rev. D {\bf 35}, 3621 (1987).}
%
\bibitem{WKB6}{R. A. Konoplya, Phys. Rev. D {\bf 68}, 124017 (2003) [arXiv:hep-th/0309030]; J. Phys. Stud. {\bf 8}, 93 (2004).}
%
%
\bibitem{Leaver}{E. W. Leaver, Proc. Roy. Soc. Lond. A {\bf 402}, 285 (1985).}
%
\bibitem{Iyer1}{S. Iyer, Phys. Rev. D {\bf 35}, 3632 (1987).}
%
\bibitem{Nollert}{H. P. Nollert, Phys. Rev. D {\bf 47}, 5253 (1993).}
%
\bibitem{Motl1}{L. Motl, Adv. Theor. Math. Phys. {\bf 6}, 1135 (2003) [arXiv:gr-qc/0212096].}
%
\bibitem{Schw2}{H. T. Cho, Phys. Rev. D {\bf 68}, 024003 (2003) [arXiv:gr-qc/ 0303078].}
%
\bibitem{Schw3}{V. Cardoso and J. P. S. Lemos, Phys. Rev. D {\bf 67}, 084020 (2003) [arXiv:gr-qc/0301078].}
%
\bibitem{Padmanabhan}{T. Padmanabhan, Classical Quantum Gravity {\bf 21}, L1 (2004) [arXiv:gr-qc/0310027].}
%
\bibitem{Cardoso}{V. Cardoso, J. P. S. Lemos, and S. Yoshida, Phys. Rev. D {\bf 69}, 044004 (2004) [arXiv:gr-qc/0309112].}
%
\bibitem{Schw.dS}{F. Mellor and I. Moss, Phys. Rev. D {\bf 41}, 403 (1990).}
%
\bibitem{Otsuki}{H. Otsuki and T. Futamase, Prog. Theor. Phys. {\bf 85}, 771 (1991).}
%
\bibitem{Moss}{I. G. Moss and J. P. Norman, Classical Quantum Gravity {\bf 19}, 2323 (2002) [arXiv:gr-qc/0201016].}
%
\bibitem{Choudhury}{T. R. Choudhury and T. Padmanabhan, Phys. Rev. D {\bf 69}, 064033 (2004) [arXiv:gr-qc/0311064].}
%
\bibitem{Schw.dS2}{V. Cardoso, J. Natario, and R. Schiappa, J. Math. Phys. (N.Y.) {\bf 45}, 4698 (2004) [arXiv:hep-th/0403132].}
%
\bibitem{KZ}{R. A. Konoplya and A. Zhidenko, J. High Energy Phys. {\bf 06} (2004) 037 [arXiv:hep-th/0402080].}
%
\bibitem{AZ}{A. Zhidenko, Classical Quantum Gravity {\bf 21}, 273 (2004) [arXiv:gr-qc/0307012].}
%
\bibitem{Cardoso2}{V. Cardoso and J. P. S. Lemos, Phys. Rev. D {\bf 64}, 084017 (2001) [arXiv:gr-qc/0105103].}
%
\bibitem{Cardoso3}{V. Cardoso, R. Konoplya, and J. P. S. Lemos, Phys. Rev. D {\bf 68}, 044024 (2003) [arXiv:gr-qc/0305037].}
%
\bibitem{RNds}{K. D. Kokkotas, and B. F. Schutz, Phys. Rev. D {\bf 37}, 3378 (1988).}
%
\bibitem{Motl2}{L. Motl and A. Neitzke, Adv. Theor. Math. Phys. {\bf 7}, 307 (2003) [arXiv:hep-th/0301173].}
%
\bibitem{KZ1}{R. A. Konoplya and A. Zhidenko, Phys. Rev. D {\bf 90}, 064048 (2014) [arXiv:1406.0019 [hep-th]].}
%
\bibitem{Kerr}{E. Berti, V. Cardoso, and S. Yoshida, Phys. Rev. D {\bf 69}, 124018 (2004) [arXiv:gr-qc/0401052].}
%
\bibitem{Musiri}{S. Musiri and G. Siopsis, Phys. Lett. B {\bf 579}, 25 (2004) [arXiv:hep-th/0309227].}
%
\bibitem{Hod}{S. Hod and U. Keshet, Classical Quantum Gravity {\bf 22}, L71 (2005) [arXiv:gr-qc/0505112].}
%
\bibitem{KN}{K. D. Kokkotas, Nuovo Cimento Soc. Ital. Fis. {\bf 108B}, 991 (1993).}
%
\bibitem{Keshet}{U. Keshet and S. Hod, Phys. Rev. D {\bf 76}, 061501 (2007) [arXiv:0705.1179 [gr-qc]].}
%
\bibitem{Zhu}{J.-M. Zhu, B. Wang, and E. Abdalla, Phys. Rev. D {\bf 63}, 124004 (2001) [arXiv:hep-th/0101133].}
%
\bibitem{Wang}{B. Wang, C.-Y. Lin, and C. Molina, Phys. Rev. D {\bf 70}, 064025 (2004) [arXiv:hep-th/0407024].}
%
\bibitem{Kodama2}{H. Kodama, R. A. Konoplya, and A. Zhidenko, Phys. Rev. D {\bf 79}, 044003 (2009) [arXiv:0812.0445 [hep-th]].}
%
\bibitem{Bronnikov}{K. A. Bronnikov, R. A. Konoplya, and A. Zhidenko, Phys. Rev. D {\bf 86}, 024028 (2012) [arXiv:1205.2224 [gr-qc]].}
%
\bibitem{Lemos}{V. Cardoso, J. P. S. Lemos, and S. Yoshida, Phys. Rev. D {\bf 70}, 124032 (2004) [arXiv:gr-qc/0410107].}
%
\bibitem{Abdalla}{E. Abdalla, B. Cuadros-Melgar, A. B. Pavan, and C. Molina, Nucl. Phys. B {\bf 752}, 40 (2006) [arXiv:gr-qc/0604033].}
%
\bibitem{Naresh}{R. Gannouji and N. Dadhich, Classical Quantum Gravity {\bf 31}, 165016 (2014) [arXiv:1311.4543 [gr-qc]].}
%
\bibitem{RBH}{A. Flachi and J. P. S. Lemos, Phys. Rev. D {\bf 87}, 024034 (2013) [arXiv:1211.6212 [gr-qc]].}
%
\bibitem{Li}{J. Li, H. Ma, and K. Lin, Phys. Rev. D {\bf 88}, 064001 (2013) [arXiv:1308.6499 [gr-qc]].}
%
\bibitem{Medved}{A. J. M. Medved, D. Martin, and M. Visser, Classical Quantum Gravity {\bf 21}, 1393 (2004) [arXiv:gr-qc/0310009]; {\bf 21}, 2393 (2004) [arXiv:gr-qc/0310097].}
%
\bibitem{Nomura}{H. Nomura and T. Tamaki, Phys. Rev. D {\bf 71}, 124033 (2005) [arXiv:hep-th/0504059].}
%
\bibitem{Fernando}{S. Fernando and J. Correa, Phys. Rev. D {\bf 86}, 064039 (2012) [arXiv:1208.5442 [gr-qc]].}
%
\bibitem{Bardeen}{J. M. Bardeen, in Proceedings of GR5, Tbilisi, USSR, 1968 (unpublished), p. 174.}
%
\bibitem{ABG}{E. Ay\'{o}n-Beato and A. Garc\'{i}a, Phys. Rev. Lett. {\bf 80}, 5056
(1998) [arXiv:gr-qc/9911046];\\ Phys. Lett. B {\bf 493}, 149
(2000) [arXiv:gr-qc/0009077].}
%
\bibitem{Bronnikov2}{K. A. Bronnikov, V. N. Melnikov, G. N. Shikin, and K. P. Staniukovich, Ann. Phys. (N.Y.) {\bf 118}, 84 (1979).}
%
\bibitem{Bronnikov3}{K. A. Bronnikov, Phys. Rev. Lett. {\bf 85}, 4641 (2000).}
%
\bibitem{Bronnikov4}{K. A. Bronnikov, Phys. Rev. D {\bf 63}, 044005 (2001) [arXiv:gr-qc/0006014].}
%
\bibitem{Moffat}{W. A. Moffat, Black holes in modified gravity (MOG), [arXiv:1412.5424 [gr-qc]].}
%
\bibitem{Hayward}{S. A. Hayward, Phys. Rev. Lett. {\bf 96}, 031103 (2006) [arXiv:gr-qc/0506126].}
%
\bibitem{Garcia}{A. Garcia, E. Hackmann, J. Kunz, C. L\"{a}mmerzahl, and A. Macias, Motion of test particles in a regular black hole space-time, [arXiv:1306.2549 [gr-qc]].}
%
\bibitem{Patil}{M. Patil and P. S. Joshi, Phys. Rev. D {\bf 86}, 044040 (2012) [arXiv:1203.1803 [gr-qc]].}
%
\bibitem{Bobir}{B. Toshmatov, B. Ahmedov, A. Abdujabbarov, and Z. Stuchl\'{i}k, Phys. Rev. D {\bf 89}, 104017 (2014) [arXiv:1404.6443 [gr-qc]].}
%
\bibitem{Stu-Sche:IJMPD:2015:}{Z. Stuchl\'{i}k and J. Schee, Int. J. Mod. Phys. D {\bf 24}, 1550020 (2015) [arXiv:1501.00015 [astro-ph.HE]].}
%
\bibitem{Stuchlik2}{J. Schee and Z. Stuchl\'{i}k, Gravitational lensing and ghost
images in the regular Bardeen no-horizon spacetimes, [arXiv:1501.00835 [astro-ph.HE]].}
%
\bibitem{SH}{Z. Stuchl\'{i}k and S. Hledik, Acta Physica Slovaca, {\bf 52}, 363 (2002).}
%
\bibitem{Pugliese1}{D. Pugliese, H. Quevedo, and R. Ruffini, Phys. Rev. D {\bf 83}, 104052 (2011) [arXiv:1103.1807 [gr-qc]].}
%
\bibitem{Pugliese2}{D. Pugliese, H. Quevedo, and R. Ruffini, Phys. Rev. D {\bf 83}, 024021 (2011) [arXiv:1012.5411 [astro-ph.HE]].}
%
\bibitem{Pugliese3}{D. Pugliese, H. Quevedo, and R. Ruffini, Circular motion in
Reissner-Nordström spacetime, [arXiv:1003.2687 [gr-qc]].}
%
\bibitem{Vieira}{R. S. S. Vieira, J. Schee, W. Kluzniak, Z. Stuchl\'{i}k, and M. Abramowicz, Phys. Rev. D {\bf 90}, 024035 (2014) [arXiv:1311.5820 [gr-qc]].}
%
\bibitem{Schee}{Z. Stuchl\'{i}k and J. Schee, Classical Quantum Gravity {\bf 31}, 195013 (2014) [arXiv:1402.2891 [astro-ph.HE]].}
%
\bibitem{Daniela}{Z. Stuchl\'{i}k and P. Slan\'{y}, Phys. Rev. D {\bf 69}, 064001 (2004) [arXiv:gr-qc/0307049].}
%
\bibitem{Stuchlik3}{Z. Stuchl\'{i}k, Bull. Astron. Inst. Czech. {\bf 31}, 129 (1980).}
%
\bibitem{Stuchlik4}{Z. Stuchl\'{i}k and J. Schee, Classical Quantum Gravity {\bf 27}, 215017 (2010) [arXiv:1101.3569 [gr-qc]].}
%
\bibitem{Stuchlik5}{Z. Stuchl\'{i}k and J. Schee, Classical Quantum Gravity {\bf 30}, 075012 (2013).}
%
\bibitem{Moreno}{C. Moreno and O. Sarbach, Phys. Rev. D {\bf 67}, 024028 (2003) [arXiv:gr-qc/0208090].}
%
\bibitem{Kimura}{M. Kimura, K. Murata, H. Ishihara, and J. Soda, Phys. Rev. D {\bf 77}, 064015 (2008) [arXiv:0712.4202 [hep-th]].}
%
\bibitem{Konoplya1}{R. A. Konoplya, Phys. Rev. D {\bf 68}, 024018 (2003) [arXiv:gr-qc/0303052].}
%
\bibitem{Konoplya12}{R. A. Konoplya, Phys. Lett. B {\bf 679}, 499 (2009) [arXiv:0905.1523 [hep-th]].}
%
\bibitem{Konoplya2}{P. Kanti and R. A. Konoplya, Phys. Rev. D {\bf 73}, 044002 (2006) [arXiv:hep-th/0512257].}
%
\bibitem{Zhidenko1}{R. A. Konoplya and A. Zhidenko, Phys. Rev. D {\bf 81}, 124036 (2010) [arXiv:1004.1284 [hep-th]]; Phys. Lett. B {\bf 686}, 199 (2010) [arXiv:0909.2138 [hep-th]].}
%


\end{thebibliography}
\end{document}